\documentstyle[aps,preprint,eqsecnum]{revtex}
\draft
\tightenlines

\makeatletter
\def\fps@figure{tbp}
\def\ftype@figure{1}
\def\ext@figure{lof}
\def\fnum@figure{\figurename\penalty10000\hskip.3em plus .1em\relax\thefigure.}
\if@floats
\def\figure{\let\@capwidth\columnwidth\@float{figure}}
\let\endfigure\end@float
\@namedef{figure*}{\let\@capwidth\textwidth\@dblfloat{figure}}
\@namedef{endfigure*}{\end@dblfloat}
\else
\def\figure{%
\let\@capwidth\columnwidth
\vskip1pc
\def\@captype{figure}%
\interlinepenalty10000 %
\@ifnextchar[{\@chuckoptarg}{}%
}%
\def\endfigure{\goodbreak\vskip1pc}%
\@namedef{figure*}{\figure}%
\@namedef{endfigure*}{\endfigure}%
\fi
\makeatother

\begin{document}
\draft
\preprint{NTZ 6/91}
\title{
Wave Functions, Evolution Equations and \\
Evolution Kernels from Light-Ray Operators of QCD}

\author{D. M\"uller, D. Robaschik, B. Geyer}
\address{
Fachbereich Physik der Universit\"at Leipzig, Germany
}
\author{F.-M. Dittes}
\address{
Institut f\"ur Kernforschung Rossendorf, Germany
}
\author{J. Ho\v{r}ej\v{s}i}
\address{
Nuclear Centre, Charles University Prague, \v{C}SFR
}
\maketitle

\begin{abstract}
The  widely  used  nonperturbative  wave  functions  and  distribution
functions  of  QCD  are  determined  as  matrix  elements of light-ray
operators.  These operators appear as large momentum limit of nonlocal
hadron  operators  or  as  summed  up  local  operators  in light-cone
expansions.  Nonforward one-particle matrix elements of such operators
lead   to   new   distribution   amplitudes  describing  both  hadrons
simultaneously.  These  distribution  functions  depend  besides other
variables   on  two  scaling  variables.  They  are  applied  for  the
description  of  exclusive  virtual  Compton scattering in the Bjorken
region  near  forward  direction and the two meson production process.
The  evolution equations for these distribution amplitudes are derived
on  the  basis of the renormalization group equation of the considered
operators.  This  includes that also the evolution kernels follow from
the   anomalous  dimensions  of  these  operators.  Relations  between
different  evolution  kernels (especially the Altarelli-Parisi and the
Brodsky-Lepage)  kernels  are  derived  and explicitly checked for the
existing  two-loop  calculations  of  QCD\@.  Technical basis of these
results  are  support  and  analytically  properties  of the anomalous
dimensions  of  light-ray  operators  obtained  with  the  help of the
$\alpha$-representation of Green's functions.
\end{abstract}
\newpage 
 
\section{Introduction}

It  is generally accepted that quantum chromodynamics (QCD) allows the
theoretical  description  of  scattering  processes  at large momentum
transfer.  Thereby,  it  is  possible  to calculate the so-called hard
scattering  subprocess  perturbatively, and for the remaining parts of
the  process  one  uses  "parton  distribution functions" \cite{AP} or
"wave   functions"  \cite{BL}.  These  functions  are  not  calculable
perturbatively,   but   have   to  be  determined  phenomenologically.
Nevertheless  additional  information  on them can be obtained because
they satisfy evolution equations.

In   this  paper  we  take  into  account  that  the  usually  applied
nonperturbative  distribution  and wave functions are directly related
to   special   matrix   elements  of   (nonlocal) light-ray operators
\cite{BL,ZC,Ge,Co}.  The  parton distribution amplitude (corresponding to
the  forward  matrix  element)  is  necessary  for  the  description of
inclusive  lepton-hadron scattering and the meson (hadron) wave functions
(corresponding  to  the  matrix element between the one-hadron and the
vacuum   state)  is  needed  for  the  treatment  of  exclusive  large
transverse  momentum  processes.  Here we discuss the question what role
plays  the  general  one-particle  matrix  element  between hadrons of
different  momenta.  To  these matrix elements belong new distribution
amplitudes    describing    both    hadrons    simultaneously.    Such
nonperturbative  distribution  amplitudes are necessary if there is no
large momentum transfer between the hadrons (otherwise we could use a
a  hard scattering part and a separate description of each hadron by a
nonperturbative  function). As an example we show that these functions
can  be  applied to the prescription of the virtual Compton scattering
amplitude  in  the  Bjorken  region  near forward scattering or to two
meson  production  processes. Thereby it is extremely interesting that
we  obtain  a  very  simple  representation  of the Compton scattering
amplitude in leading order. In fact, the virtual Compton amplitude can
be  represented  as  a  convolution of a hard scattering part with the
distribution  amplitude,  so  that  the  analogy  to the simple parton
picture remains.

A  second  topic  of  this paper is the investigation of the evolution
equations and evolution kernels for these nonperturbative distribution
amplitudes.  Here  we  take into account that all these amplitudes are
matrix  elements  of  light-ray  operators, satisfying renormalization
group  equations.  Therefore,  the  evolution  equations are basically
renormalization group equations. Of course, by forming matrix elements
the  renormalization group equations for light-ray operators turn over
to evolution equations for these matrix elements. Using such a general
procedure,  we  are  able (in principle) to derive evolution equations
for all matrix elements of such operators. The Brodsky-Lepage equation
\cite{BL}  for  the  meson  wave  function  and  the  Altarelli-Parisi
equation  \cite{AP}  for  the  quark  distribution  function  of  deep
inelastic scattering appear as special cases. It is important that the
evolution  kernels  are  functionals of the anomalous dimension of the
considered  operator.  A trivial but interesting point is that thereby
the  range  of  the  variables  may  change  according  to the support
properties  of the distribution amplitude. A very general kernel which
is  in  fact the complete anomalous dimension of the underlying 
light-ray operator is the extended Brodsky-Lepage (BL) kernel. It 
represents a  partly  diagonalized kernel generating all one-parametric 
evolution equations. In the one-loop approximation, a complete 
diagonalized form of  such a kernel following from nonlocal 
(conformal) operator product expansions has been studied in Ref. \cite{BB}.

Besides   of   these  general  considerations  we  discuss  also  some
interesting  technical  problems concerning the explicit determination
of  these kernels. In QCD, the BL-kernel is calculated as an evolution
kernel  in  a  restricted  range  of  the  variables.  The interesting
question  is: Exists there a possibility to determine the extended 
BL-kernel  knowing  this restricted kernel only? We prove that the partly
Fourier transformed BL-kernel has holomorphic properties allowing such
an  extension.  In the following these results are applied to the one-
and   two-loop   calculations  of  the  BL-kernel  in  QCD\@.  Without
additional  diagrammatic  calculations  we  are  able to determine the
extended  BL-kernel  \cite{DG}  of  QCD\@.  Finally,  we perform the
limiting   procedure  leading  from  the  extended  BL-kernel  to  the
Altarelli-Parisi  (AP)  kernel,  and  compare the resulting expression
with  the  directly calculated results \cite{2A,2D,2B}. In this way we
obtain   a   consistency  check  for  the  very  complicated  two-loop
calculations  \cite{2D,2B}. Taking into consideration the recent check
\cite{DM} based on conformal Wardidentities and consistence equations,
we conclude that the BL-kernel has been correctly computed.

All  our  investigations  concern  the  flavour  nonsinglet operators.
Investigations of singlet operators are in principle possible, however
technically   more   complicated.

It remains to give some remarks concerning the light-ray operators and
its anomalous dimensions. Light-ray operators were first introduced by
O. I. Zavialov \cite{AZ} in connection with his proof of the light-cone
operator  product  expansion.  In opposite to the standard formulation
\cite{OP}  he  did  not  use the expansion in terms of local operators
\cite{OP}  but introduced light-ray operators which are in fact summed
up  local  operators.  We also use this technique for the treatment of
the  virtual  Compton  scattering  amplitude.  We hope to convince the
reader  that  this  method  is very effective. An independent approach
based  on  the same ideas was developed and applied from the Leningrad
group  \cite{BB,BAOP}.  Another  source for the introduction of light-ray
operators   are   the  speculations  about  hadron  operators  in  QCD
\cite{HAOP}.  Hadrons  are  composed  of  quarks and gluons, therefore
hadron  operators  have to be bi- or trilocal operators containing the
basic fields of QCD. The postulate of gauge invariance leads to string
operators  finally.  If  we  want  to  apply  these  operators in hard
scattering  processes  then for large internal momenta these operators
turn  over to the light-ray operators considered here. It is now clear
that  for  example  bilocal  operators where the distance between both
positions  of  the  involved  field  operators  is  light-like possess
additional  divergences. Therefore they need a special renormalization
procedure  which  leads  to  more  complicated anomalous dimensions as
usually discussed.

To  avoid  complicated  notations  in the main text we generally write
down  renormalized  expressions.  Only  in  the  Appendix  we also use
unrenormalized   expressions.  An  important  technical  task  is  the
investigation   of   the  anomalous  dimensions  of  these  operators,
especially  the  correct  determination  of  its domain of definition.
Using  the  $\alpha$-representation  of Feynman graphs, we first study
this  problem  for  the  scalar theory and afterwards for QCD in axial
gauge.  For  the  last  case  we modify the $\alpha$-representation to
include  the  more  complicated  gluon propagator of QCD in light-cone
gauge.

The  paper is organized as follows: In the second section we introduce
the  generalized  distribution amplitudes as matrix elements of light-ray
operators.  In  the  next  section  we apply these distribution
amplitudes  for  the  prescription  of  the  exclusive virtual Compton
scattering  and  the  two  meson  production by photons. In the fourth
section  we  derive  the  evolution  equations  for  these amplitudes.
Problems  concerning explicit expressions (two-loop calculations) and
relation  between different evolution kernels are discussed in the last
section.   Important   technical  details  concerning  the  domain  of
definition of the anomalous dimension, the perturbative calculation of
hard  scattering  parts  and other problems are shifted to Appendices.
The  Appendices  are needed for the proofs only, the reader uninterested in
technical details could skip them.

\section{Nonperturbative Inputs: Distribution Amplitudes}

Up  to  now  the  application of nonperturbatively determined "parton
distribution functions" and "wave functions" is an unavoidable part of
perturbative  QCD  calculations.  We start our considerations with the
introduction  of more general distribution amplitudes corresponding to
nonforward scattering processes.

In  the  QCD  picture  hadrons are built up from quarks and gluons. So
possible  hadron operators are nonlocal operators containing quark and
gluon  fields. Typical meson operators are composed of quark
and antiquark operators
\begin{eqnarray}
\label{DOp}
O^a (x_1, x_2) =\; :\!\bar \psi(x_1)  \Gamma \lambda^a U(x_1, x_2)
\psi(x_2)\!:.
\end{eqnarray}
The path ordered phase factor,
\begin{eqnarray}
 U(x_1,x_2)=
 {\cal P}\exp{\left\{-ig \int_{x_2}^{x_1} dx_\mu  A^\mu (x) \right\}}
\end{eqnarray}
ensures  the  gauge  invariance  of  the  considered operator. $A^{\mu
}=A^{\mu }_c t^c$ denotes the gluon field, $t^c$ are the generators of
the  colour  group,  $\psi$ is the quark field, $\Gamma$ symbolizes the
necessary spin structure and $\lambda^a$ is a generator of the flavour
group  corresponding  to the considered meson. At large internal quark
momenta  the operator (\ref{DOp}) turns over to an asymptotic operator
defined on a straightline light-like path
\begin{eqnarray}
\label{DOV}
O^a (\kappa_+,\kappa_- ;\tilde n) =
	\; :\!\bar \psi(\kappa_1 \tilde  n)  (\tilde  n\gamma)
	\lambda^a U(\kappa_1 \tilde n,\kappa_2 \tilde n)
	\psi(\kappa_2 \tilde n)\!:,\quad \kappa_\pm =(\kappa_2 \pm \kappa_1)/2,
\end{eqnarray}
where
\begin{eqnarray} 
 U(\kappa_1\tilde n,\kappa_2\tilde  n)=
  {\cal P} \exp{\left\{-ig \int_{\kappa_2}^{\kappa_1} d\tau  \tilde nA(\tau
\tilde n) \right\}}.
\nonumber
\end{eqnarray}
The vector $\tilde
n$,  with  ${\tilde  n}^2  = 0$, defines the light-ray pointing into the
direction of the large momentum flow of this process.

For the description of mesons involved in hard QCD processes one needs an
"asymptotic" wave function as nonperturbative input. This
meson wave function $\Phi^a (x,Q^2)$ depending on the distribution
parameter  $x$  $(0  \leq  x \leq 1)$ and the momentum transfer $Q^2$
can be defined as the
expectation  value  of  a  nonlocal  (light-ray) operator lying on the
light-cone \cite{BL,ZC,Ge}:
\FL
\begin{eqnarray}
\label{DMF}
\Phi^a (x =(1+t)/2,Q^2)=\int{d\kappa_- |\tilde n P| \over 2\pi}
e^{i\kappa_-(\tilde nP)t} {1\over \tilde nP}
<0|O^a(\kappa_-;\tilde n)|P>|\scriptstyle {\mu}^2=Q^2 \displaystyle,
\end{eqnarray}
where 
\FL
\begin{eqnarray}
\label{DO}
O^a (\kappa_-;\tilde n)=O^a (\kappa_+=0,\kappa_-;\tilde n)=
\; :\!\bar \psi(-\kappa_- \tilde  n)  (\tilde  n
\gamma) \lambda^a U(-\kappa_- \tilde n,\kappa_- \tilde n) \psi(\kappa_-
\tilde n)\!:.
\end{eqnarray}
Here  $|P>$  denotes  the  one-particle  state  of  a  scalar meson of
momentum  $P$  and $O^a(\kappa_- ;\tilde n)$ is the light-ray operator
with  the same flavour content. As renormalization point we choose the
typical  large  momentum  $Q$  of  the basic process to which the wave
function  contributes.The  factor  $1/(\tilde  n P)$ was introduced to
compensate  the $\tilde n$-dependence of the factor $\tilde n \gamma $
in  the operator $O^a(\kappa_- ;\tilde n)$. Contrary to the definition
in  Ref. \cite{BL}  our  expression  (\ref{DMF})  is gauge invariant and 
Lorentz-covariant.   In   an   infinite   momentum  frame  with  
$P=(|P_z|,0,0,P_z)$
(neglecting the masses of the particles) and $\tilde n=(1,0,0,-1)$ the
distribution  parameter  $x$  can be interpreted as usually: $xP$ is the
fraction of the meson momentum $P$ which is carried by the quark. Then
the  mesonwave  function  is  the  probability amplitude for finding a
quark-antiquark  pair  in  the  meson  in  dependence  of the momentum
fraction parameter $x$ and the momentum transfer $ Q^2$.

As next, we define the quark distribution functions $q_i(z,Q^2)$ \cite{AP}
with the distribution parameter $z$. These functions are necessary for
the description of deep inelastic scattering. At first such
functions have been introduced phenomenologically, later on as Mellin
transforms  of  moments  of forward matrix elements of local operators
\cite{AP}. If we take into account that the operator
\begin{eqnarray}
O_i (\kappa_+ ,\kappa_-)= \; : \! \bar \psi_i(\kappa_1 \tilde n) (\tilde n
\gamma ) U(\kappa_1 \tilde n,\kappa_2 \tilde n) \psi_i(\kappa_2 \tilde n)
\!:
\end{eqnarray}
is the "generating operator" for all local operators,
\begin{eqnarray}
 ((\tilde  n   D)^{n_1}\bar\psi_i(0)) (\tilde n\gamma) (\tilde n D)^{n_2}
 \psi_i(0) = {\partial^{n_1} \over \partial \kappa_1^{n_1}}
{\partial^{n_2}\over\partial\kappa_2^{n_2}} 
O_i (\kappa_+ ,\kappa_-)|\scriptstyle \kappa_1=\kappa_2=0 \displaystyle ,
\end{eqnarray}
then  a  definition  of  the  quark  distribution  function  as Fourier
transform of the one-particle matrix element of the light-ray operator
\cite{Co},
\begin{eqnarray}
O_i (\kappa_- ;\tilde n)= \; : \! \bar \psi_i(-\kappa_- \tilde n) (\tilde n
\gamma ) U(-\kappa_- \tilde n,\kappa_- \tilde n) \psi_i(\kappa_- \tilde n)
\! :,
\end{eqnarray}
seems to be understandable.
Here $\psi_i$ denotes the quark field with flavour component $i$.
However, group theoretically this operator has no
definite flavour content which leads to mixing problems with the flavour
singlet gluon operator. To avoid this problem we consider here the flavour
nonsinglet distribution function only:
\begin{eqnarray}
\label{DQF}
q^a (z,Q^2)= \int     {d\kappa_- |\tilde n P| \over 2\pi }
e^{2i\kappa_- (\tilde n P)z}{1\over\tilde n P} <P|O^a (\kappa_- ;\tilde n)
|P>|\scriptstyle {\mu}^2=Q^2 \displaystyle.
\end{eqnarray}

Physically  this  distribution function describes the probability
of finding internal  quarks  with  the  momentum  fraction $ zP$, 
where $P$ is the external hadron momentum.
If  the matrix $\lambda^a$ is choosen diagonal, then
for positive distribution parameter $z$ this function represents a linear
combination of the quark distribution functions, and the extension
for $z<0$  can be interpreted as a
linear combination of the antiquark distribution functions.

Both  functions  in (\ref{DMF}) and (\ref{DQF}) are expectation values
of  the  same operator used, however, for different physical processes
and having different interpretations.

If  we  look  at  the definition (\ref{DQF}) of the quark distribution
function  given  as Fourier transform of a forward matrix element of a
light-ray  operator  then  the  question  arises: What is the physical
meaning  of  nonforward matrix elements? As an example we study in the
next   section   two-photon  processes  in  the  Bjorken  region.
The  idea is that the product of the two electromagnetic currents near
light-like  distances  can  be  expressed  with  help  of  coefficient
functions and the light-ray operators $O^a (\kappa_- ;\tilde x)$. (This
generalizes the standard local operator product expansion).

If  we consider the virtual Compton scattering amplitude in nonforward
direction,  than  we  have  to  form nonforward matrix elements of these
light-ray   operators (see the next section).   This  leads  us  directly
to  the  following distribution   amplitude,
\FL
\begin{eqnarray}  \label{DIF}  q^a  (t,\tau
,\mu^2)=  \int {d\kappa_- |\tilde n P_+| \over 2\pi } e^{i\kappa_-
(\tilde  n  P_+)t}{1\over\tilde  nP_+  }<P_2|O^a  (\kappa_- ;\tilde n)
|P_1>|\scriptstyle  {\tilde  n P}_-=\tau {\tilde n P}_+ \displaystyle,
\end{eqnarray}
with $P_\pm = P_2 \pm P_1$.
This  function  depends  on  the  distribution  parameter  $t$ and the
quotient $\tau = {\tilde n P}_- / {\tilde n P}_+$ of the projection of
momenta $P_\pm $ onto  a  light-like direction $\tilde n$, on the
renormalization point $\mu$ and the scalar products of the external
momenta $P_i P_j$. For  physical states $|P_1>$ and $|P_2>$ the
additional variable $\tau$ is restricted by
\begin{eqnarray}
 |\tau| = \left |{{\tilde nP_-} \over {\tilde nP_+}} \right | =\left |
{{P_-^0- P_-^{\scriptscriptstyle \| \displaystyle}} \over {P_+^0-
P_+^{\scriptscriptstyle \| \displaystyle}}}\right| \leq 1,\quad
P^{\scriptscriptstyle \| \displaystyle} = {{\vec{\tilde n}\vec
P}\over{|\vec{\tilde n}|}}.
\end{eqnarray}

With  the  help  of  the  $\alpha$-representation it is possible to
investigate  the  support  properties  of  the  function  $q^a (t,\tau
,\mu^2)$  with  respect to the variable $t$. Extending the proof given
in Ref. \cite{Ge} for $\tau =1$ to arbitrary values of $\tau$, it turns out
that  
\begin{eqnarray}  
\label{212}
q^a  (t,\tau ,Q^2) = 0 \qquad \hbox{for} \quad
|t| > 1. 
\end{eqnarray}

The  distribution  amplitude  $q^a (t,\tau ,Q^2)$ is especially suited
for  the description of nonforward processes near to the forward case.
Namely,  if  there  is  no  large  momentum  transfer  between the two
hadrons,  then  it is impossible to describe the hadrons by individual
wave  functions  and  a  hard scattering part, i.e. the parton picture
with  two  separate  wave functions cannot be applied. Here one has to
apply  one  function  describing  both  hadrons.  An example of such a
process  will be discussed in the next section (it is interesting that
a similar function was already proposed by Ref. \cite{Chase}).

On  the  other  hand,  it  is  possible  to  consider  the  nonforward
distribution   amplitude  (\ref{DIF})  as  an  interpolating  function
between  the quark distribution function (\ref{DQF}) (for $P_1 = P_2$)
and  the  meson  wave  function  (\ref{DMF}),  i.e.  in the limit $P_2
\rightarrow  0$ too. Of course, the state obtained by $P_2 \rightarrow
0$  is  not  the  vacuum  state,  but  for mathematical considerations
concerning  the  connection  of  evolution  equations  for forward and
nonforward  processes  this  is very useful. As result in section V we
obtain  relations  between  different  evolution  kernels  (especially
between  the  Brodsky-Lepage  kernel  and the Altarelli-Parisi kernel)
which are highly nontrivial.

Therefore, we  can  perform  two limits
\begin{eqnarray}
\Phi^a(x=(1+t)/2,Q^2)&=&   \lim_{\tau   \to   -1}q^a  (t,\tau  ,Q^2)  \quad
\mbox{ meson   wave  function  (formally!),}  
\nonumber \\  
q^a  (z,Q^2)&=&\lim_{\tau  \to  0} q^a (z,\tau ,Q^2) \quad 
\mbox{ quark distribution function (formally!)}.
\nonumber\\
\end{eqnarray}

In short we add some further properties of the distribution amplitude
(\ref{DIF}):
\begin{itemize}
\item Because of the conjugation properties of the light-ray operators
$\left( O^a (\kappa_- ;\tilde n) \right)^+ = O^a (-\kappa_- ;\tilde n)$ it
satisfies
\begin{eqnarray}
\left(q^a(t,\tau ,\mu^2)\right )^* = q^a (t,-\tau ,\mu^2).
\end{eqnarray}
This means that this amplitude is real for $ P_1 = P_2 $ .
\item The normalization of this distribution amplitude follows  
from the definition (\ref{DIF})  and (\ref{DO}), 
\begin{eqnarray}
\label{Nor}
\int dt q^a (t,\tau ,\mu^2) = (\tilde nP_+)^{-1} \tilde
n^\nu<P_2|J_\nu^a(0)|P_1>|\scriptstyle {\tilde n P}_-=\tau
{\tilde n P}_+ \displaystyle ,
\end{eqnarray}
where $J_\nu^a(0) =\; :\!\bar \psi(0) \gamma_\nu \lambda^a
\psi(0)\!:$ is a current with the flavour content $a$.

\end{itemize}

At  last  we  give  a  representation  of the generalized distribution
amplitude  with the help of a "spectral" function which is helpful for
later calculations. Taking into account the ideas of the general 
Jost-Lehmann  representation for $T$-amplitudes of two local operators (our
operators  (\ref{DO})  are  in  fact  bilocal if we use the light-cone
gauge  in  QCD) then the spectral functions have a finite support with
respect to the new distribution variables $z_+$ and $z_-$. In this way
the  matrix  elements  of  light-ray  operators  can  be  expressed by
"spectral" functions $f^a(z_+,z_-,\mu^2)$ as follows
\begin{eqnarray}
\label{FT}
   {1\over \tilde  nP_+}   <P_2|O^a(\kappa_- ;\tilde n)|P_1>   =   
\int\int  dz_+dz_-  
e^{-i\kappa_-(\tilde      nP_-)z_-     -i\kappa_-(\tilde     nP_+)z_+}
f^a(z_+,z_-,\mu^2).
\nonumber\\
\end{eqnarray}
If we insert now this equation (\ref{FT}) into the definition of the
distribution function (\ref{DIF}), we obtain a new representation
\begin{eqnarray}
\label{FTIF}
            q^a(t,\tau ,Q^2)&=&\int\int     dz_+dz_-
                           \int    {d\kappa_-|\tilde nP_+|\over 2\pi}
          e^{i\kappa_-(\tilde nP_+)(t-z_+-\tau z_-)} f^a(z_+,z_-,Q^2)
\nonumber\\
                 &=& \int     dz_- f^a(z_+=t-\tau z_-,z_-,Q^2),
\end{eqnarray}
which  shows  in which way the mathematically independent distribution
variables $z_+$ and $z_-$ turn over to the more physical variables $t$
and $\tau$.

\section{Structure Functions for Two-Photon Processes}

One   of   the   first  crucial  test  of  perturbative  QCD  was  the
investigation  of the deep inelastic electron proton scattering in the
Bjorken   region.  Theoretically  it  implies  the  knowledge  of  the
imaginary  part of the virtual Compton scattering amplitude. The first
QCD  treatments  of  this  process  relied  on the light-cone operator
product  expansion  and  its  relation to the moments of the structure
functions \cite{TI}.

Another   important   class  of  processes  are  exclusive  scattering
processes   of   ha\-drons   at  large  momentum  transfer  \cite{BL}.
Perturbative   QCD   calculations  demand  the  introduction  of  wave
functions   describing   the   incoming   and  outgoing  hadrons.  The
perturbation  theory  itself  is  applicable to a hard scattering part
which connects these wave functions. The justification of perturbation
theory  is  related  to  the presence of large momenta. Whereas in the
case   of  deep  inelastic  scattering  the  scattering  amplitude  is
dominated  by  the  contributions from the light-cone singularities in
the  $x$-space, here it must not be the case. Therefore its perturbative
treatment  in QCD can be quite different and of course the kinematical
regions of both processes are also different.

Here  we want to investigate the complete Compton scattering amplitude
in  the  Bjorken  region  near forward scattering \cite{WM}. Note that
usually  perturbative  considerations  of exclusive Compton scattering
amplitudes are performed in the fixed angle region - were experimental
data  are  present  or  expected \cite{KN}. If we consider the Compton
amplitude   in   a   generalized   Bjorken  region  then  experimental
investigations with present possibilities are very hard to realize. On
the  other hand from a theoretical point of view it embeds the Compton
amplitude  for  forward  scattering  and  is  therefore interesting in
itself. Because of the smallness of the momentum transfer between both
hadrons  it  is  not  possible to introduce separately nonperturbative
amplitudes  (or  wave  functions)  for  each  hadron. Surprisingly the
treatment  of  this  process  leads  in  a  very  natural  way  to the
introduction  of  new  distribution  amplitudes which for the limiting
case  of  forward  scattering  turn  over  into  the well-known parton
distribution functions of deep inelastic scattering. Furthermore these
new  distribution amplitudes satisfy new evolution equations \cite{DG}
which  turn  over  to  the Altarelli-Parisi equation \cite{AP} for the
case of forward scattering.

To show the virtues of this construction we study additionally the two
meson production by two photons \cite{AM}. For small momentum transfer
between  both mesons also here we introduce one distribution amplitude
describing  both  mesons.

\subsection{Exclusive Virtual Compton Scattering}

Here  we  consider the virtual nonforward Compton scattering amplitude
of a spinless particle near forward direction but at a large off-shell
momentum of the the incoming photon:

\begin{eqnarray}
\label{ScA}
T_{\mu\nu}(P_+,P_-,q) =
   i \int d^4 x\, e^{iqx}\,<P_2|T\left(J_\mu\left(\frac{x}{2}\right)
                         J_\nu\left(\frac{-x}{2}\right)\right)|P_1>,
\end{eqnarray}
where
$J_\mu(x) = (1/2):\overline{\psi}(x) \gamma_\mu
            (\lambda^3 - \lambda^8/\sqrt 3)\psi(x) : $
is the electromagnetic current of the hadrons (for flavour $SU(3)$).

The  considered kinematics defines a generalized Bjorken region where,
in  close  analogy  to  deep  inelastic  scattering,  this  process is
dominated by contributions from the light-cone.
\begin{eqnarray}
  \gamma^\ast (q_1) + \hbox{H}(P_1) = \gamma^\ast(q_2) + \hbox{H}(P_2),
\end{eqnarray}
\begin{eqnarray}
  P_+ &=& P_1 +P_2 = (2E,\vec 0),\qquad  E = \sqrt{(m^2 + \vec p^2)} ,
\nonumber\\
  P_- &=& P_2 - P_1 = q_1 - q_2= (0, -2\vec p),\quad
  q = (1/2) (q_1 +q_2).
\end{eqnarray}
The  last  expressions  inside  the  brackets denote the values of the
momenta  in  the  Breit  frame,  respectively. The generalized Bjorken
region is given by
\begin{eqnarray}
\nu = P_+q = 2 E q_0\rightarrow \infty, \qquad
Q^2 = -q^2 \rightarrow \infty, 
\end{eqnarray}
with the scaling variables
\begin{eqnarray}
   \xi  &=&  {-q^2  \over P_+q} , \qquad \eta = {P_- q \over P_+q} = {q_1^2
   - q_2^2 \over 2\nu} .
\end{eqnarray}

To  understand  this process better we introduce the angle between the
vectors $\vec p$  and  $\vec q$ in the Breit frame by
$\cos\phi = {\vec p \vec q \over |\vec p| |\vec q|}$. 
In terms of these variables we get
\begin{eqnarray}
   q^2_1 \approx -\left( \xi -{|\vec p|\over E} \cos\phi \right) \nu,\quad
   q^2_2 \approx -\left( \xi +{|\vec p|\over E} \cos\phi \right)\nu,\quad
    \eta \approx {|\vec p|\over E} \cos\phi .
\end{eqnarray}
In  opposite  to  deep  inelastic scattering the variable $\xi$ is not
restricted  to  $  0 \le \xi \le 1$; for example $ q^2_2 =0$ demands $
\xi \approx -(|\vec p|/E)\cos\phi$.

In Appendix A  we show that in this region the helicity amplitudes
$T(\lambda',\lambda) = \varepsilon_2^\mu(\lambda')
    T_{\mu\nu} \varepsilon_1^\nu(\lambda)$
asymptotically are given by
\begin{eqnarray}
\label{AScA}
T(\lambda',\lambda) =
	(1/2)\varepsilon_2^\nu(\lambda')\varepsilon_{1\nu}(\lambda)
	T_\mu^{\mbox{\ }\mu}
\end{eqnarray} 
for  the  transverse helicities, and vanish otherwise. Therefore, only
the trace of the scattering amplitude has to be considered.

The  first treatment of the Compton scattering amplitude with the help
of a (local) light-cone
operator product expansion \cite{OP,TI}, symbolized by
\begin{eqnarray*}
J(x)J(0) =
\sum_{n=0}^{\infty} C_n(x^2) x^{\mu_1}...x^{\mu_n} O_{\mu_1...\mu_n}(0),
\end{eqnarray*}
has   the   serious   disadvantage   that  the  amplitude  has  to  be
reconstructed  from  an  infinite  sum.  This may be accepted for deep
inelastic  scattering  where the famous connection between the moments
of  the  structure  functions  and the expectation values of the local
light-cone  operators exists, but in general it is unsatisfactory. The
application of the nonlocal light-cone expansion \cite{Ge,BB,AZ,Br,Za}
is  a  suitable  possibility to overcome this drawback. In our special
case it reads (in leading order):
\begin{eqnarray}
\label{LCE}
J^\mu\left(\frac{x}{2}\right)  J_\mu\left(\frac{-x}{2}\right)  \approx
\int   d\kappa_+   d\kappa_-  \,  F_a(x^2,\kappa_+,\kappa_-;\mu^2)  \,
O^a(\kappa_+,\kappa_-;\tilde{x})_{(\mu^2)}
\end{eqnarray}
with  the light-ray    operators   given   in   (\ref{DOV}).
The light-like vector
\begin{eqnarray*}
\tilde  x(x,\rho)=  x  +  \rho {x\rho \over \rho^2} \left(\sqrt{1-{x^2
\rho^2 \over(x\rho )^2}}-1\right)
\end{eqnarray*}
is  determined  by  $x$  and  parameterized by a fixed constant vector
$\rho$.
The singular coefficient functions $F_a$ are determined
perturbatively; in the Born approximation they are given by
\begin{eqnarray}
F_a(x^2,\kappa_+,\kappa_-) \approx i e_a
\left(2\pi^2(x^2 - i\epsilon)^2\right)^{-1}\delta(\kappa_+)
\left(\delta(\kappa_- -1/2) - \delta(\kappa_- + 1/2) \right),
\end{eqnarray}
with 
$e_a =(2/9)\delta_{a0} + (1/6)\delta_{a3} + (1/6\sqrt 3)\delta_{a8}$.
Possible contact terms arising at the point $x=0$ lead to trivial 
contributions to the real part of the scattering amplitude and are suppressed.

Inserting  the  light-cone expansion (\ref{LCE}) into the trace of the
scattering   amplitude   (\ref{ScA})   and  using  the  representation
(\ref{FT}) for the matrix elements $<P_2|O^a(\kappa_-;\tilde x)|P_1>$,
\begin{eqnarray*}
         <P_2|O^a(\kappa_-;\tilde x)|P_1>=\tilde xP_+
                                           \int \int     dz_+dz_- 
           e^{-i\kappa_-(\tilde xP_-)z_- -i\kappa_-(\tilde xP_+)z_+}
                       f^a(z_+,z_-;P_iP_j,\mu^2),
\end{eqnarray*}
we get
\begin{eqnarray}
         T_\mu^{\mbox{\ }\mu}(P_+,P_- ,q) \approx
                   2\int\int    dz_+dz_- 
\tilde F^a(\xi ,\eta ;z_+,z_-) f^a(z_+,z_-;P_iP_j,\mu^2=Q^2),
\end{eqnarray}
with
\begin{eqnarray*}
    \tilde F^a =-e^a \int    {d^4x\over 2\pi^2}\int     d\kappa_- 
        e^{ix[q-(P_+z_+ +P_- z_-)\kappa_-]}{P_+x\over (x^2 -i\epsilon )^2}
\left(\delta (\kappa_--1/2)-\delta (\kappa_-+1/2)\right) .
\end{eqnarray*}
(Here the approximation $\tilde x \approx x$ has been used.)

To  obtain a useful expression in momentum space the Fourier transform
has to be carried out. Using
\begin{eqnarray}
     \int    {d^4x\over 2\pi^2} e^{ix[q-k]}{P_+x\over (x^2-i\epsilon )^2}
    ={P_+(q-k)\over (q-k)^2 +i\epsilon} \approx
		{P_+q \over q^2-2qk +i\epsilon},
                            \quad k=(P_+z_++P_- z_-)\kappa_-
\nonumber\\
\end{eqnarray}
we find
\begin{eqnarray}
      T_\mu^{\mbox{\ } \mu}\approx 2\int\int     dz_+dz_- 
     \left({1\over \xi +z_++\eta z_-}-{1\over\xi - z_+ -\eta z_-}\right)
                              e^af^a(z_+,z_-;P_iP_j,\mu^2=Q^2).
\nonumber\\
\end{eqnarray}

In  this  form the result is unsatisfactory. It depends explicitely on
the  two  distribution  variables  $z_+,\  z_-$  of  the distribution
amplitude  $f^a(z_+,z_-)$  .  However, it is possible to simplify this
expression  further, and to derive an expression for the $T$-amplitude
containing only one integral over the standard distribution amplitude.
For this purpose we substitute $z_+=t-\eta z_-$ so that
\begin{eqnarray}
\label{ZScA}
       T_\mu ^{\mbox{\ } \mu} &\approx& 2 \int     dt
                     \left({1\over\xi +t}-{1\over\xi -t}\right)
           \int     dz_- e^a f^a(z_+=t-\eta z_-,z_-;P_iP_j ,\mu^2 = Q^2)
\nonumber\\
                              &\approx& 2 \int     dt 
                     \left({1\over\xi +t}-{1\over\xi -t}\right)
             e^a q^a(t,\eta ;P_iP_j ;\mu^2=Q^2),
\end{eqnarray}
where we have used the definition (\ref{FTIF}) with $\tau$ replaced by
$\eta$,
\begin{eqnarray*}
                 q^a(t,\eta ;P_iP_j,Q^2)=
      \int     dz_-  f^a(z_+=t-\eta z_-,z_-;P_iP_j ,Q^2).
\end{eqnarray*}
Taking into account Eq. (\ref{AScA}), we obtain the final result
\begin{eqnarray}
\label{EScA}
                T(\lambda ',\lambda)\approx
         \varepsilon_2^\mu (\lambda ')\varepsilon_{1\mu} (\lambda)
        \int     dt \left({1\over\xi +t}-{1\over\xi -t}\right)
                 e^a q^a(t,\eta ;P_iP_j ;\mu^2=Q^2).
\end{eqnarray}

Note that in order to obtain Eq. (\ref{EScA}) it was necessary to turn
over  from  the  nonperturbative  function  $f^a(z_+,z_-)$ back to the
distribution amplitude $q^a(t,\eta )$, however, with changed variables.
The  original  definition  (\ref{DIF})  of $q^a(t,\tau )$ contains the
light-like  vector $\tilde n$ which would be given here by $\tilde x$.
This  would lead to another definition of the variable $\tau$ by $\tau
=\tilde  xP_- /\tilde xP_+$ which does not make sense. But instead of
this  unphysical  variable  here the physical variable $\eta = qP_-
/qP_+$  appears.  This  amplitude  generalizes  the  parton distribution
function  of  deep  inelastic  scattering  with  the  above  mentioned
properties. Note that $q^a(t,Q^2) = q^a(t,\eta =0,Q^2)$ is essentially
the parton distribution function.

According to the optical theorem the absorptive part of the scattering
amplitude $ W_{\mu\nu} =\hbox{Im}T_{\mu\nu}/(2\pi) $ is related to the
total  cross  section  in deep inelastic scattering. Since $q^a(t,\eta
=0,Q^2)$ is real from equation (\ref{ZScA}) we get directly
\begin{eqnarray}
  W_\mu^{\mbox{\ }\mu} &\approx& -\int_{-1}^{1} dt (\delta( \xi - t) 
                                                   - \delta(\xi +t))
                  e_aq^a(t,\eta=0;\mu^2 =Q^2)
\nonumber\\
     &\approx& -\left(e_aq^a(\xi,\eta=0;\mu^2 =Q^2)
               - e_aq^a(-\xi,\eta=0;\mu^2 =Q^2)\right).
\end{eqnarray}
Here
\begin{eqnarray}
     -q^a(-\xi,\eta=0;\mu^2 =Q^2)=\bar q^a(\xi,\eta=0;\mu^2 =Q^2)
\end{eqnarray}
represents the antiquark distribution amplitude.

\subsection{Two Meson Production}

As  a  further instructive example we discuss the two meson production
by two photons
\begin{eqnarray}
\gamma^\ast (q_1) + \gamma^\ast (q_2) = \hbox{M}(P_1) + \hbox{M}(P_2).
\end{eqnarray}
This  is  a  crossed  process  with  respect  to  the  virtual Compton
scattering. As variables we choose
\begin{eqnarray}
       q =(1/2) (q_1 - q_2),\quad P_+= P_1 + P_2,\quad P_- =P_2 - P_1.
\end{eqnarray}
Again the generalized Bjorken region is defined by
\begin{eqnarray}
\nu = P_+q = 2 E q_0\rightarrow \infty, \qquad
Q^2 = -q^2 \rightarrow \infty, 
\end{eqnarray}
with the scaling variables
\begin{eqnarray}
   \xi  &=&  {-q^2  \over P_+q} , \qquad \eta = {P_- q \over P_+q}.
\end{eqnarray}

The  virtual  photons  are  taken  at  large  off shell momenta $q_1^2
\rightarrow -\infty, q_2^2 \rightarrow -\infty$. Then the range of the
scaling  variable  is $|\xi| \ge 1$. In the center of mass system with
fixed  energy  of  the  two  mesons  these  scaling  variables  can be
expressed as
\begin{eqnarray}
   \xi = {-q^2 \over P_+q} = -{q_1^2 + q_2^2 \over q_1^2 - q_2^2} ,\quad
   \eta = {P_- q \over P_+q} \approx {\vec p \vec q \over E q_0} 
        \approx \cos\phi.
\end{eqnarray}

In the leading approximation the scattering amplitude can be
written as [see (\ref{AScA})]
\FL
\begin{eqnarray}
T(P_+,P_-,q;\lambda',\lambda) ={1\over 2} 
\varepsilon_2^\nu (\lambda') \varepsilon_{1\nu}(\lambda)
   i \int d^4 x\, e^{iqx}\,<P_1,P_2|T\left(J_\mu\left(\frac{x}{2}\right)
                                  J^\mu\left(\frac{-x}{2}\right)\right)  |0>.
\nonumber\\
\end{eqnarray}
After  the  application  of  the  operator  product expansion, and the
calculation  of the coefficient function we arrive at the same type of
expressions as (\ref{ZScA})
\begin{eqnarray}
\label{MPA}
T^{\mbox{\ }\mu}_\mu(P_+,P_-,q) \approx 2 \int dt
\left(\frac{1}{\xi +t} -\frac{1}{\xi - t}\right)
e_aq^a_m(t,\eta;\mu^2 =Q^2),
\end{eqnarray}
where the distribution amplitudes reads now
\begin{eqnarray}
q^a_m(t,\eta;\mu^2) =
\int \frac{d\kappa_-|\tilde{x}P_+|}{2\pi(\tilde{x}P_+)}
e^{i\kappa_-(\tilde{x}P_+)t}
<P_1,P_2| O^a(\kappa_-;\tilde{x})_{\mu^2}|0>_{\big|\tilde{x}P_-
=\eta\tilde{x}P_+}.
\end{eqnarray}
It  is  interesting  that  a similar distribution function was already
introduced  by  Chase \cite{Chase} in the two photon
production of two jets.
In  the limit $\eta \rightarrow 1$ we obtain the known result for the
production of a scalar meson by two photons for large $Q^2$ \cite{Ge,GM}.

Note  that  also  here  we  generalize  the  meson  wave function to a
two-meson  wave function in a straightforward way. Of course we cannot
predict   the  dependence  of  this  wave  function  on  its  internal
parameters  (this  includes  also its $P_1P_2$ dependence). This is in
contrast  to  the  large  momentum  transfer calculation where $P_1P_2
\rightarrow  \infty$.  Here one introduces for each meson its own wave
function  and  then  it  is  clear that the $P_1P_2$ dependence of the
process is included into the hard scattering part.

\pagebreak[1]
\subsection{Parton interpretation}

Here we try to interpret the foregoing results.

There  are  two  remarkable  differences of the foregoing processes in
comparison   with   the  usually  discussed  forward  or  fixed  angle
scattering processes:
\begin{enumerate}
\item There appear generalized distribution amplitudes describing both
    hadrons simultaneously. These  distribution functions   depend,
    besides other variables, also on the scaling variable $\eta$ defined
	above.
\item The leading contribution of the hard scattering part consists of an
   interaction  of  both  photons with one and the same internal quark line.
    The momenta of the other quarks are not forced to be large, so that they 
    are confined and can be described implicitely by
    the generalized distribution amplitude. Fig. \ref{FIG1} illustrates
    this remark for the case of the virtual Compton scattering.
    A  standard  treatment  of  exclusive  pion  (or  nucleon)  Compton
    scattering at fixed angles would include much more diagrams.
\end{enumerate}
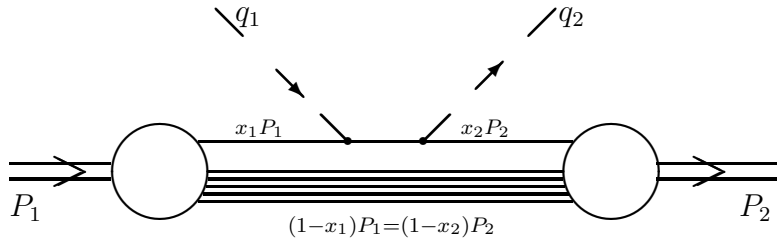
\begin{figure}[h]
\unitlength1cm
\begin{picture}(12,5)(-2,0)
\thicklines
\put(0,1.9){\line(1,0){1.35}}
\put(0,2.1){\line(1,0){1.35}}
\put(1,2){\line(-2,-1){0.4}}
\put(1,2){\line(-2,1){0.4}}
\put(0,1.4){$P_1$}
\put(3,2.5){$\scriptstyle x_1P_1$}
\put(2,2){\circle{1.3}}

\put(2.51,2.4){\line(1,0){4.98}}
\put(2.65,2){\line(1,0){4.7}}
\put(2.64,1.9){\line(1,0){4.715}}
\put(2.62,1.8){\line(1,0){4.76}}
\put(2.58,1.7){\line(1,0){4.85}}
\put(2.51,1.6){\line(1,0){4.98}}
\put(3.7,1.2){$\scriptstyle (1-x_1)P_1=(1-x_2)P_2 $}

\multiput(4.5,2.4)(-1.4,1.4){2}{\line(-1,1){0.36}}
\put(3,4){$q_1$}
\put(3.55,3.35){\vector(1,-1){0.36}}
\multiput(5.5,2.4)(1.4,1.4){2}{\line(1,1){0.36}}
\put(7.3,4){$q_2$}
\put(6.2,3.1){\vector(1,1){0.36}}
\put(4.5,2.4){\circle*{0.1}}
\put(5.5,2.4){\circle*{0.1}}

\put(6,2.5){$\scriptstyle x_2P_2 $}
\put(8,2){\circle{1.3}}
\put(8.6,1.9){\line(1,0){1.6}}
\put(8.6,2.1){\line(1,0){1.6}}
\put(9.5,2){\line(-2,-1){0.4}}
\put(9.5,2){\line(-2,1){0.4}}
\put(9.7,1.4){$P_2$}
\end{picture}
\caption[FIG.1]{Leading contribution for the virtual Compton scattering 
                 in the Bjorken region.}
\label{FIG1}
\end{figure}

\begin{figure}[h]
\unitlength1cm
\begin{picture}(12,5)(-2,0)
\thicklines

\put(0,1.9){\line(1,0){1.35}}
\put(0,2.1){\line(1,0){1.35}}
\put(1,2){\line(-2,-1){0.4}}
\put(1,2){\line(-2,1){0.4}}
\put(0,1.4){$P_1$}
\put(2,2){\circle{1.3}}

\put(2.55,2.4){\line(1,0){4.9}}
\put(2.55,1.6){\line(1,0){4.9}}
\multiput(4,2.4)(-1.4,1.4){2}{\line(-1,1){0.36}}
\put(2.5,4){$q_1$}
\put(3.05,3.35){\vector(1,-1){0.36}}
\multiput(5,1.6)(0.7,0.7){2}{\line(1,1){0.36}}
\multiput(7.1,3.7)(0.7,0.7){1}{\line(1,1){0.36}}
\put(7.5,4){$q_2$}
\put(6.4,3){\vector(1,1){0.36}}
\put(4,2.4){\circle*{0.1}}
\put(5,1.6){\circle*{0.1}}
\put(4.5,2.4){\circle*{0.1}}
\put(4.5,1.6){\circle*{0.1}}
\multiput(4.5,1.7)(0,0.4){2}{\oval(0.2,0.2)[l]}
\multiput(4.5,1.9)(0,0.4){2}{\oval(0.2,0.2)[r]}

\put(8,2){\circle{1.3}}
\put(8.6,1.9){\line(1,0){1.6}}
\put(8.6,2.1){\line(1,0){1.6}}
\put(9.5,2){\line(-2,-1){0.4}}
\put(9.5,2){\line(-2,1){0.4}}
\put(9.7,1.4){$P_2$}
\end{picture}
\caption[FIG.2]{Leading contribution for the virtual Compton amplitude at
fixed angles.}
\label{FIG2}
\end{figure}
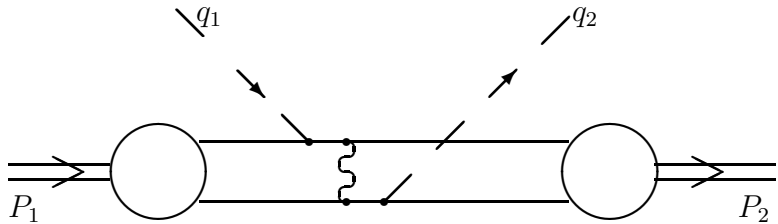

It  remains  an
interpretation  of  $t$  in the parton language as momentum fractions of
the external  hadrons. In the case of the exclusive  virtual  Compton
scattering we obtain
\begin{eqnarray}
      t = {x_1 P_1 + x_2 P_2 \over P_1 + P_2}
\end{eqnarray}
in an infinite momentum frame with respect to $P_1+P_2$. This variable
is a natural generalization of the earlier introduced distribution 
parameters $z$ or $x$ of the parton distribution function and the meson wave
function, respectively.
We  underline that the standard calculation of the pion Compton effect
at fixed angles for example uses diagrams of Fig. \ref{FIG2}.
\noindent
Here  in  opposite  to  the  foregoing  case  the  photons act on both
internal  lines.  Because  of the large momentum transfer between both
hadrons  it  is possible to introduce separate wave functions for each
hadron.  So  the  $P_1P_2$  dependence  is  a natural part of the hard
scattering  subprocess.  If  in addition we can go over to the Bjorken
region,  then  both formalisms should be applicable and should lead to
the same result.

\section{Derivation of the General Evolution Equation}

In  this section, we derive an evolution equation for the distribution
amplitude   (\ref{DIF})   which   contains   as   limiting  cases  the
Brodsky-Lepage  and  the  Altarelli-Parisi  equations.  This evolution
equation  exploits the renormalization group equation of the light-ray
operators   (\ref{DO}).   For   this   reason  we  have  to  know  the
renormalization  properties  of  these  operators and especially their
anomalous  dimensions.  Afterwards  we  can  apply  these  results  to
expectation  values  of  these  operators  which  are contained in the
distribution amplitude.

It is now well-known that bilocal operators where the distance between
both  positions  of the involved field operators is light-like possess
additional  divergences. Therefore they need a special renormalization
procedure.  In  the  scalar  case  the  renormalization  of  light-ray
operators  has  been  studied  from  a  general point of view by S. A.
Anikin  and  O. I. Zavialov \cite{AZ}. These operators need   special
Z-factors   depending   on  the  parameters  of  these  operators  and
additional anomalous dimensions \cite{Bo}.
A general thorough proof for the renormalization of these operators in
QCD  with  covariant  gauge  fixing  we do not know. However practical
calculations of the Z-factors respectively the corresponding anomalous
dimensions  exist  at  one loop \cite{Ge} and two loops for restricted
variable  ranges \cite{2D,2B} for covariant gauge fixing as well as in
the light-cone gauge.

It turns out that for a straightforward mathematical treatment the set
of  operators  (\ref{DO})  is  not well suited. We use, therefore, the
more general operators (\ref{DOV})
\begin{eqnarray}
O^a(\kappa_+ ,\kappa_- ;\tilde n) = \; :\! \overline{\psi }(\kappa_1 \tilde
n) (\tilde n \gamma) \lambda^a U(\kappa_1 \tilde n,\kappa_2 \tilde n)
\psi(\kappa_2 \tilde n)\!:, \quad \kappa_\pm = (\kappa_2 \pm \kappa_1 )/2.
\end{eqnarray}
These  operators differ from the original ones by a translation. Their
renormalization group equation reads formally
\begin{eqnarray}
\mu {d \over d\mu}O^a (\kappa_+,\kappa_-;\tilde n)_{(\mu^2)}  &=& 
\int d\kappa'_+ d\kappa'_- 
\big[ \gamma (\kappa_+,\kappa_-,\kappa'_+,\kappa'_-;\alpha_s (\mu^2 ) ) 
 - 2\gamma_\psi(\alpha_s (\mu^2 ))
\nonumber\\
&&\quad \qquad\times 
\delta(\kappa_+ -\kappa'_+)  \delta (\kappa_+ - \kappa'_+)\big] O^a
(\kappa'_+,\kappa'_-;\tilde n)_{(\mu^2)}.
\end{eqnarray}
This  integral equation contains two integration variables. Using 
symmetry properties a more
simple form of this equation will be derived in
Appendix B \cite{Br}, namely
\begin{eqnarray}
\label{RGE}
\mu {d \over d\mu}O^a (\kappa_+,\kappa_-;\tilde n)_{(\mu^2)}  =
\int && dw_+ dw_-
\big[ \gamma (w_+,w_-;\alpha_s(\mu^2)) - 2\gamma_\psi(\alpha_s(\mu^2)) 
\nonumber\\
&&\times\delta(w_+) \delta(1-w_-) \big] 
O^a(w_+\kappa_-+\kappa_+,w_-\kappa_-;\tilde n)_{(\mu^2)}.
\nonumber\\
\end{eqnarray}

In  the  Appendix  B  we also investigate the anomalous dimensions $\gamma
(w_+,w_-)$  of  the  above defined operators. As result, we obtain the
following support restriction (\ref{BSR})
\begin{eqnarray}
\gamma (w_+,w_-) \neq 0\qquad \hbox{for}\quad |w_\pm |\leq 1,\mbox{} |w_+\pm
w_-| \leq 1. 
\end{eqnarray}
Additionally, from the transformation properties of
the operators  under charge conjugation it follows
$\gamma (w_+,w_-) = \gamma (-w_+,w_-)$ [Eq. (\ref{BCC})].

Technically  we  investigate  in  Appendix B the divergent part of all
one-particle-irreducible  (1PI)  diagrams  with  the  insertion of one
light-ray operator for scalar field theory and QCD in light-cone gauge.
As  essential  tool  we use the $\alpha$-representation. However we do
not prove or carry out the renormalization procedure of the considered
operator.  For  the  investigation of the support properties of the
Z-factors  and the anomalous dimensions of these operators this is not
necessary.  So  we  study the support properties of the divergent term
with  respect  to  the  parameters  $\kappa_+$  and  $\kappa_-$ of the
inserted  operator  and the parameters $ \kappa'_+$ and $\kappa'_-$ of
the  appearing  counter terms. This allows us to determine the support
properties of the anomalous dimension of the light-ray operators.

After  these  preliminaries we turn to the derivation of the evolution
equation.  For  this purpose we differentiate the general distribution
amplitude  $q^a(t,\tau  ,\mu^2)$  with  respect to the renormalization
parameter  $\mu$.  Thereby  we  take  into  account its representation
(\ref{DIF}) in terms of matrix elements of the light-ray operator. The
differentiation of this operator can be performed with the help of its
renormalization   group   equation   (\ref{RGE}).   A  straightforward
calculation runs as follows:
\FL
\begin{eqnarray}
\label{AbEvE1}
            \mu {d\over d\mu} q^a(t,\tau ,\mu^2)&=&
    \int {d\kappa_-|\tilde nP_+|\over 2\pi (\tilde nP_+)}
                         e^{i\kappa_-(\tilde nP_+)t} \mu{d\over d\mu}
                         <P_2|O^a(\kappa_+=0,\kappa_-;\tilde n)|P_1>
         |\scriptstyle \tilde nP_- = \tau (\tilde nP_+) \displaystyle 
\nonumber\\
&=&\int    dw_- dw_+
\int    {d\kappa_-|\tilde nP_+|\over 2\pi (\tilde nP_+)}
e^{i\kappa_-(\tilde nP_+)t} \big[\gamma (w_+,w_-) -
2\gamma_\psi
\nonumber \\
&& \qquad \times \delta (w_+)\delta (1-w_-)\big]
<P_2|O^a(\kappa_-w_+,\kappa_-w_-;\tilde n)|P_1>|\scriptstyle \tilde nP_-
=\tau (\tilde nP_+)\displaystyle . 
\nonumber\\
\end{eqnarray}
As next step we exploit the translation invariance of the matrix element 
\begin{eqnarray}
<P_2|O^a(\kappa_-w_+,\kappa_-w_-;\tilde n)|P_1>=
<P_2|O^a(\kappa_+=0,\kappa_-w_-;\tilde n)|P_1>e^{i\kappa_-w_+(\tilde nP_-)}
\end{eqnarray}
and introduce the new variables  $\kappa'_-=\kappa_-w_-$ and  $t' =
(t+\tau w_+)/w_-$ into Eq. (\ref{AbEvE1}):
\begin{eqnarray}
\mu {d\over d\mu}q^a(t,\tau ,\mu^2)&=&\int {d^2\underline{w}\over |w_-|}
\int     {d\kappa '_-|\tilde nP_+| \over 2\pi(\tilde nP_+)}
e^{i(t+\tau w_+)\kappa'_-(\tilde nP_+)/w_-} \big[\gamma (w_+,w_-)
-2\gamma_\psi
\nonumber\\
&&\qquad \times  
\delta (w_+) \delta (1-w_-)\big] 
<P_2|O^a(\kappa'_+=0,\kappa'_-;\tilde n)|P_1>|\scriptstyle \tilde nP_- =
\tau (\tilde nP_+) \displaystyle 
\nonumber \\
&=&\int    dt' \int    {d^2\underline{w}\over |w_-|} \delta 
\left(t'-{{t+\tau  w_+} \over w_-}\right) \left[\gamma (w_+,w_-)
-2\gamma_\psi \delta (w_+) \delta (1-w_-)\right] 
\nonumber\\
& &\qquad\qquad\qquad\qquad\qquad\times q^a(t',\tau , \mu^2).
\end{eqnarray}
As result, we obtain the evolution equation
\begin{eqnarray}
\label{EIF}
Q^2{d\over dQ^2}q^a(t,\tau ,Q^2)=\int_{-1}^1 {dt'\over |2\tau|}
\left(\gamma \left({t\over \tau},{t'\over \tau}\right)-2\gamma_\psi \delta
\left({t\over \tau} -{t'\over \tau}\right)\right) q^a(t',\tau ,Q^2)
\end{eqnarray}
 with the evolution kernel
\begin{eqnarray}
\label{DEvK}
\gamma (t,t') = \int    dw_- \gamma (w_+=t'w_--t,w_-).
\end{eqnarray}

From the renormalization group invariance of $\int dt q^a(t,\tau ,
\mu^2)$ (see Eq. (\ref{Nor})) it follows
\begin{eqnarray}
\int    dt \gamma (t,t') = 2\gamma_\psi,
\end{eqnarray}
so that the standard "+"-definition for the generalized function
\begin{eqnarray}
\label{RP}
[\gamma (t,t')]_+=\gamma(t,t')-\delta(t-t')\int    dt'' 
\gamma (t'',t')=\gamma (t,t')-2\gamma_\psi \delta (t-t')
\end{eqnarray}
arises  in  a  very  natural way. We see that instead of the anomalous
dimension  there  appears  an  evolution  kernel  which  contains  the
anomalous  dimension  as an essential input. As it becomes clear later
on    it  seems  to  be  natural to speak of our general evolution
kernel  as  of  an  extended  BL-kernel.  The  reason  for this is the
following:   Restricting   $\gamma(t,t')$   to  the  parameter  region
$|t|,|t'|\leq  1$,  it  coincides  with  the  evolution kernel for the
hadron   wave  function,  but  Eq.  (\ref{EIF})  provides  us  with  a
meaningfull  definition  also  outside of this region.

In  this  way,  we  obtain  evolution  equations  for  all forward and
nonforward  matrix elements. To our knowledge, in the literature up to
now  only  two  examples  have  been investigated: the case of forward
scattering  and  the case of the meson wave function. Here, both cases
are  contained as limits $P_2 \to P_1$ or $P_2 \to 0$ (see section II).
We write these limits down in a condensed form:
\begin{itemize}
\item Evolution
equation for the quark distribution function (forward scattering):
\begin{eqnarray} q^a (z,Q^2)&=& \lim_{\tau \to 0} q^a (z,\tau ,Q^2),
\nonumber \\ Q^2{d\over dQ^2}q^a(z,Q^2)&=&\int_{-1}^1 {dz'\over |z'|}
P\left({z\over z'};\alpha_s(Q^2)\right) q^a(z',Q^2),
\end{eqnarray}
\begin{eqnarray}
\label{APK}
|z'|^{-1}P\left({z\over z'}\right)&=&\lim_{\tau \to 0} {1\over |2\tau |}
\left[\gamma\left({z\over \tau},{z'\over \tau}\right)\right]_+, \quad
\mbox{Altarelli-Parisi-(Lipatov) kernel.}
\nonumber\\
\end{eqnarray}
\item 
Evolution  equation  for  hadron  wave  functions:\\  This case can be
obtained  as  a  formal  limit  too [compare the remarks after
Eq. (\ref{212})].  
\begin{eqnarray}
\Phi^a (x=(1+t)/2,Q^2)&=& \lim_{\tau \to -1}q^a (t,\tau ,Q^2), \nonumber\\
Q^2{d\over dQ^2} \Phi^a(x,Q^2)&=&\int_{0}^1 dy
V_{BL}(x,y;\alpha_s(Q^2)) \Phi^a(y,Q^2),
\end{eqnarray}
\begin{eqnarray}
\label{BLK}
V_{BL}(x,y)&=&[\gamma (2x-1,2y-1)]_+|\scriptstyle 0\leq x,y\leq 1
\displaystyle , \quad \mbox{Brodsky-Lepage kernel.}
\end{eqnarray}
Here, the variables $x=(1+t)/2$, $y = (1+t')/2$ are  restricted    to
$0\leq x,y\leq 1$.
\end{itemize}
Note  that  both  kernels  (\ref{APK})  and (\ref{BLK}) are calculated
separately  in  QCD\@.  From  the  above considerations it is obvious,
however,  that  they  have  a  common  origin,  namely  the  anomalous
dimension  $\gamma(w_+,w_-)$  of  the light-ray operators being hidden
within  the  corresponding  amplitudes.  We  return  to  this point in
section V.

Let  us  add a remark concerning the distribution amplitude introduced
for  the  nonforward  virtual  Compton  scattering.  This distribution
amplitude  appearing  in  (\ref{EScA}) or (\ref{MPA}) satisfies in the
flavour nonsinglet case the evolution equation
\begin{eqnarray}
               Q^2{d\over d Q^2} q^a(t,\eta ,Q^2)=
                                   \int_{-1}^1{dt'\over |2\eta |}
       \left[\gamma\left({t\over\eta},{t'\over\eta} \right)\right]_+
                             q^a(t',\eta ,Q^2).
\end{eqnarray}
Here,  the  distribution  function  depends  on variables with a clear
physical  interpretation.  It  coincides  formally with the definition
(\ref{EIF})  so  that  no  further proof is needed. Nevertheless, this
equation can be proved directly: It is possible to derive an evolution
equation   for   $f^a(z_+,z_-,Q^2)$   starting   from   the   original
renormalization  group  equation,  and to use Eq. (\ref{FTIF}) for the
amplitude  $q^a(t',\eta ,Q^2)$. This chain of relations is independent
of  the  concrete  physical  meaning  of  the  variables,  so that the
evolution equation (\ref{EIF}) is valid for the changed variables too.
In  the  singlet  case  we  have  a  mixing  problem  with  a  gluonic
contribution.  This  problem  could  be  solved  in  a straightforward
manner, too.

\section{The Extended BL-Kernel, Relations \\ between the BL- and AP-Kernels}
 
In section IV we derived the general evolution kernel (an extended
BL-kernel) for distribution amplitudes containing the (restricted)
BL-kernel  and  the  AP-kernel  as special cases. In QCD the BL-kernel is
calculated  as  an  evolution  kernel  for  a  restricted range of the
variables.  The  interesting questions are: Exists there a possibility
to  determine  the  extended  BL-kernel knowing this restricted kernel
only?  Furthermore, if such a procedure exists, is it then possible to
apply  it  to the existing two-loop QCD calculations? If yes, then one
should  be  able  to  perform the limit to the Altarelli-Parisi kernel
which  should then coincide with the already existing results obtained
by  straightforward  computation.  We  solve  these  problems  in  the
following three subsections.

\subsection{General Aspects}

Here, we  discuss the general solution of the problems stated above.
Starting from the support properties of the anomalous dimension
$\gamma (w_+,w_-)$ and from the definition of the evolution kernel
(\ref{DEvK}), we study its domain of definition. Afterwards,  we
study the extension procedure. It turns out that the restricted BL-kernel
contains already the essential information and that an explicit
continuation procedure can be prescribed.

The first question  is: What is the correct region in the $(t,t')$-plane
where the kernel $\gamma (t,t')$ is defined in fact. To answer it,
we start from the general representation (\ref{DEvK}) of this kernel
\begin{eqnarray*}
          \gamma (t,t')=\int\int     dw_-dw_+
               \delta (w_+-t+t'w_-)\gamma (w_+,w_-)
\end{eqnarray*}
and use all known symmetry properties and support restrictions:
\begin{eqnarray*}
           \gamma (w_+,w_-)=\gamma (-w_+,w_-), \qquad
           |w_\pm |\leq 1,\quad |w_+\pm w_-|\leq 1.
\end{eqnarray*}
After some algebra we obtain the following representation in the
$(t,t')$-plane \cite{DG}
\begin{eqnarray}
\label{GAMTRE}
 \gamma(t,t')&=&
      [\theta (t-t')\theta (1-t)-\theta (t'-t)\theta (t-1)] f(t,t') 
\nonumber\\
   &&+[\theta (t'-t)\theta (1+t)-\theta (t-t')\theta (-t-1)] f(-t,-t')
  \nonumber\\
   &&+[\theta (-t-t')\theta (1+t)-\theta (t'+t)\theta (-t-1)] g(-t,t')
  \nonumber\\
   &&+[\theta (t+t')\theta (1-t)-\theta (-t'-t)\theta (t-1)] g(t,-t'),
  \end{eqnarray}
where the functions $f(t,t')$ and $g(t,t')$ are given by
\begin{eqnarray}
\label{INTRE}
f(t,t')&=&\int_0^{1-t\over 1-t'}dw_- \gamma (w_+=t-w_- t',w_-),
\nonumber\\
g(-t,t')&=&\int_0^{1+t\over 1-t'}dw_- \gamma (w_+= -t-w_- t',-w_-).
\end{eqnarray}
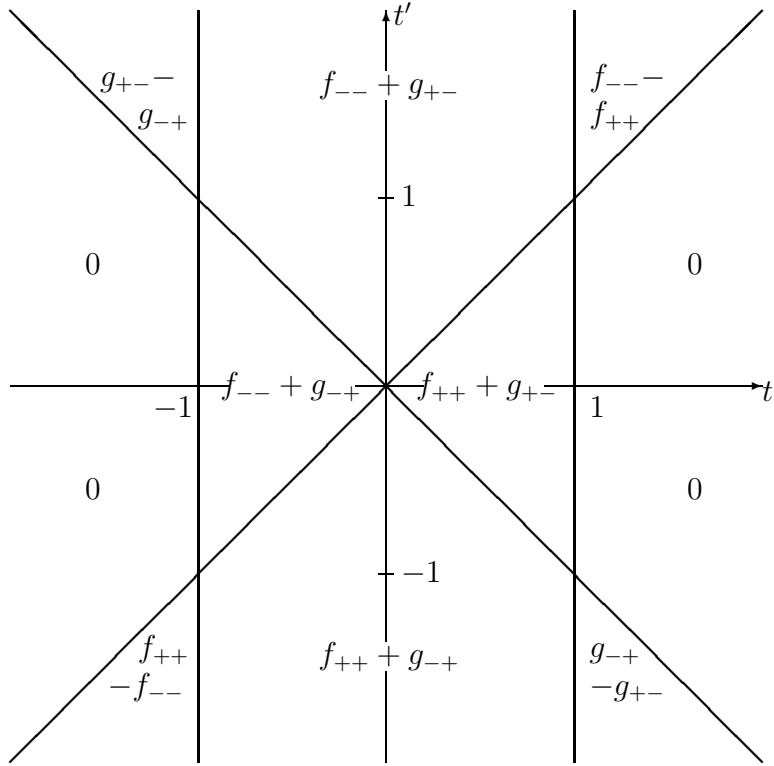
\begin{figure}[h]
\unitlength1cm
\begin{picture}(10,11)(-2,-0.5)
 
\put(0,5){\line(1,0){2.9}}    \put(4.6,5){\line(1,0){0.9}}
\put(7.1,5){\vector(1,0){2.9}}
\put(2.5,4.9){\line(0,1){0.2}}
\put(7.5,4.9){\line(0,1){0.2}}
\put(10,4.8){$t$}
\put(1.9,4.6){$-1$}
\put(7.7,4.6){$1$}
 
\put(5,0){\line(0,1){1.2}} \put(5,1.6){\line(0,1){7.2}}
\put(5,9.2){\vector(0,1){0.8}}
\put(4.9,2.5){\line(1,0){0.2}}
\put(4.9,7.5){\line(1,0){0.2}}
\put(5.1,9.8){$t'$}
\put(5.2,2.4){$-1$}
\put(5.2,7.4){$1$}
 
\thicklines
\multiput(2.5,0)(5,0){2}{\line(0,10){10}}
 
\put(0,0){\line(1,1){10}}
\put(10,0){\line(-1,1){10}}
 
\put(1.2,9){$g_{+-}-$}    \put(1.7,8.5){$g_{-+}$}
\put(7.7,9){$f_{--}-$}    \put(7.7,8.5){$f_{++}$}
\put(1.7,1.4){$f_{++}$}    \put(1.3,0.9){$-f_{--}$}
\put(7.7,1.4){$g_{-+}$}     \put(7.7,0.9){$-g_{+-}$}
 
\put(4.1,8.9){$f_{--}+g_{+-}$}
\put(2.8,4.9){$f_{--}+g_{-+}$}
\put(5.4,4.9){$f_{++}+g_{+-}$}
\put(4.1,1.3){$f_{++}+g_{-+}$}
 
\put(1,6.5){$0$}
\put(9,6.5){$0$}
\put(1,3.5){$0$}
\put(9,3.5){$0$}
\end{picture}
\caption[FIG.3]{Support  of  $\gamma  (t,t')$, where
$f_{\pm\pm}=f(\pm t,\pm t')$,  and  $g_{\pm\mp}=g(\pm   t,\mp   t')$
are   defined   by Eqs. (\ref{INTRE}).}
\label{FIG3}
\end{figure}
\noindent
Note, that for $|t|,|t'|>1$ because of support restrictions the lower
boundary in the integral representations (\ref{INTRE}) is not attained.
However, in these regions  only the differences
\begin{eqnarray*}
         f(t,t')-f(-t,-t')\quad \hbox{and} \quad g(-t,t')-g(t,-t')
\end{eqnarray*}
appear in Eq. (\ref{GAMTRE}),
so that such undetermined contributions eliminate each other. It is
clear, therefore,  that the evolution kernel is defined in the complete
$(t,t')$-plane as shown in Fig. \ref{FIG3}.

The second question is now: Assuming we are able to calculate the anomalous
dimension in the region $-1\leq t,t'\leq 1,$ is it possible to determine
it in the whole $(t,t')$-plane? This problem contains the
principal question whether the restricted BL-kernel contains already the
full information on the general anomalous dimension. For solving this
problem we perform a Fourier transform of the  extended BL-kernel and
later on of the restricted BL-kernel. It will be shown that because of the
holomorphic properties of these functions it is possible to identify them.
 
So we perform a Fourier transform of the extended BL-kernel. Thereby, we
take into account the known support restrictions explicitly
\begin{eqnarray}
                      \tilde\gamma(\lambda ,t')&=&
            \int     dt e^{i\lambda t}[\gamma(t,t')]_+
\nonumber\\
&=&
  \int\!\!\!\!\!\!\!\!\!\!\!\!\!\int\limits_{{|w_+|,|w_- |\leq 1\atop
    |w_+\pm w_-|\leq 1} \mbox{\ \ \ }}\!\!\!\!\!\!\!\!\!\!\!\!\!dw_+dw_- 
\gamma (w_+,w_-)\left(e^{i\lambda (w_+ +t'w_-)}-e^{i\lambda t'}\right).
\end{eqnarray}
Because of the finite integration region in the Fourier integral with
respect to the variables $w_+,\ w_-$ the resulting Fourier transform of the
anomalous dimension is an entire function of the variables $\lambda$  
and $t'$ \cite{GS}.
 
Now, we define a second entire function of the same variables by the
Fourier transform of the restricted BL-kernel
\begin{eqnarray}
                      \tilde V(\lambda ,t')&=&
              \int_{-1}^1 dt e^{i\lambda t} [V(t,t')]_+
\nonumber\\
&=&  \int\!\!\!\!\!\!\!\!\!\!\!\!\!\int\limits_{{|w_+|,|w_- |\leq 1\atop
    |w_+\pm w_-|\leq 1}\mbox{\ \ \ }}\!\!\!\!\!\!\!\!\!\!\!\!\!dw_+dw_- 
             \theta (1-w_- t'-w_+ ) \theta (w_- t'+w_+ +1)
\nonumber\\& &\qquad\qquad\times
  \gamma (w_+,w_-)\left(e^{i\lambda (w_+ +t'w_- )}-e^{i\lambda t'} \right).
\end{eqnarray}
Both functions have to be compared. The essential point is  that, due to
 the known support properties,  for $|t'|\leq 1$ it follows automatically
also $|t|\leq 1$ so that for $|t'|\leq 1$ there exists a region where both
expression $\tilde\gamma (\lambda ,t')$ and $\tilde V(\lambda ,t')$ are
identical.  Therefore,  knowing the Fourier transform of the BL-kernel
we can obtain the Fourier transform of the extended BL-kernel by analytic
continuation. After performing the reversed transformation
\begin{eqnarray}
     [\gamma(t,t')]_+=\int{d\lambda\over 2\pi } e^{-i\lambda t} \{\tilde
                       V(\lambda ,t')\}_{AC}
\end{eqnarray}
we obtain $\gamma (t,t')$ completely.
$\{\tilde V(\lambda ,t')\}_{AC}$ denotes formally the analytic continuation
procedure to be carried out.
 
Consequently, it is sufficient to perform all diagrammatic calculations in
the restricted region $|t|,\ |t'|\leq 1$.
The essential information on the anomalous dimension is already contained
inside of this region.

\subsection{The Extended BL-Kernel of QCD in Two-Loop Approximation}

As an exercise to the foregoing investigations we will extend the
Brodsky-Lepage kernel of QCD into the complete $(t,t')$-plane
 starting from  the explicit one- and
two-loop  calculations of the restricted BL-kernel  \cite{2D,2B}.

The  known  calculations  of  the   evolution   kernel   \cite{2D,2B}
are performed in the range $-1\leq t,t'\leq1$,  which
was sufficient for studying the evolution of the hadron wave function.  The
difficult two-loop  calculation have been done in the variables  $x  =
(1+t)/2,\ y = (1+t')/2$. So for the rest of the paper  we shall  prefer
these variables too. The result of the above mentioned calculations is
\begin{eqnarray}
\label{56}
      V(x,y)&=&{\alpha_s\over 2\pi}c_F[V_0(x,y)]_++
                \left({\alpha_s \over 2\pi}\right)^2
                 c_F[c_FV_F(x,y)+(c_G/2)V_G(x,y)
\nonumber\\
             &&\hspace{6cm}\mbox{}+(N_F/2)V_N(x,y)]_++O(\alpha_s^3),
\end{eqnarray}
with
\begin{mathletters}
\begin{eqnarray}
\label{57a}
                 V_0(x,y)= \theta (y-x) F(x,y)+
        \left\{{x\to\overline{x}\atop y\to\overline{y}}\right\},
\end{eqnarray}
\begin{eqnarray}
\label{57b}
                 V_N (x,y) = \theta (y-x) v_N (x,y) +
         \left\{{x\to\overline{x} \atop y \to \overline{y}}\right\},
\end{eqnarray}
\begin{eqnarray}
\label{57c}
            V_G (x,y) = \theta (y-x) v_G (x,y) + G(x,y) +
        \left\{{x\to \overline{x} \atop y \to \overline{y}}\right\},
\end{eqnarray}
\begin{eqnarray}
\label{57d}
    V_F (x,y)=\theta (y-x) v_F (x,y)-{x\over 2\overline{y}}k(x)-G(x,y) +
        \left\{{x \to \overline{x} \atop y \to \overline{y}}\right\}.
\end{eqnarray}
\end{mathletters}
Here, the following  auxiliary functions are used:
\begin{mathletters}
\begin{eqnarray}
\label{58a}
           F(x,y) = {x\over y}\left(1-{1\over x-y}\right),
\end{eqnarray}
\begin{eqnarray}
\label{58b} 
           v_N(x,y)=-{10\over 9}F-{2\over 3}{x\over y}-
                      {2\over 3}F\ln\left({x\over y}\right),
\end{eqnarray}
\begin{eqnarray}
\label{58c}
             v_G(x,y)={67\over 9}F+{17\over 3}{x\over y}+
                      {11\over 3}F \ln \left({x\over y}\right),
\end{eqnarray}
\begin{eqnarray}
\label{58d}
         v_F(x,y)&=& -{\pi^2\over 3}F+{x\over y}-
\left({3\over 2}F-{x\over 2\overline{y}}\right)\ln \left({x\over y}\right)-
 (F-\overline{F}) \ln \left({x\over y}\right)
\nonumber\\
&&\hspace{3cm}\times\ln\left(1-{x\over y}\right)
+\left(F+{x\over 2\overline{y}}\right) \ln^2\left({x\over y}\right)
\nonumber \\
                &=& -{\pi^2\over 3}F+{x\over y}-
                          {3\over 2}F\ln\left({x\over y}\right)-
                        \left(F-{1\over x-y}\right)\ln\left({x\over y}\right)
\nonumber \\
                 &&\hspace{1.7cm}\times\ln\left(1-{x\over y}\right)
+F\ln^2 \left({x\over y}\right)+{x\over 2\overline{y}}k\left({x\over y}\right),
\end{eqnarray}
\begin{eqnarray}
\label{58e}
            k(x) = \ln (x)(1 + \ln (x) - 2\ln (\overline{x})),
\end{eqnarray}
\begin{eqnarray}
\label{Gxy}
 G(x,y)&=&2\theta (y-x)[\overline{F} \hbox{Li}_2(\overline{x})+
          \overline{F}\ln(y)\ln(\overline{x})-F\hbox{Li}_2(\overline{y})]+
          \theta (x+y-1)\Bigg[2(F-\overline{F})
\nonumber\\
          &&\times\hbox{Li}_2\left(1-{x\over y}\right)+
          (F-\overline{F}) \ln^2 (y)-
          2F\ln(y)\ln(x)+2F\hbox{Li}_2(\overline{y})-
          2F\hbox{Li}_2(x)\Bigg],
\nonumber\\
\end{eqnarray}
\end{mathletters}
where the Spence function Li$_2$ is defined by
\begin{eqnarray}
           \hbox{Li}_2(x)=-\int_0^x dz {\ln(1-z)\over z}.
\end{eqnarray}
As shorthand notation we have introduced
\begin{eqnarray}
            \overline{x}=1-x,\quad \overline{y}=1-y,
            \quad\overline{F}\equiv\overline{F}(x,y) = F(1-x,1-y).
\end{eqnarray}
The group theoretical constants are $c_F=4/3,\ c_G=3$. $N_F$ denotes the
number of quark flavors. Contrary to the usual notation of the
kernel in Ref. \cite{2D,2B} we have changed the notation by
\begin{eqnarray*}
       G(x,y)+2\theta (y-x)\overline{F}\ln(y)\ln(\overline{x}) \to G(x,y).
\end{eqnarray*}
Furthermore, for convenience we have introduced the function $k(x)$, and in
the last row of (\ref{58d}) we have used the decomposition
$\overline{F}(x,y)=-{x\over \overline{y}}+{1\over x-y}$.

The obtained kernel $V(x,y)$ is not defined for $x=y$, however the
"+"-definition provides a regularization prescription.
 
The kernel (\ref{56}) has to be extended using the foregoing results. Let us
begin with the results of the one-loop calculations (\ref{57a}) and
(\ref{58a}). According to our procedure, we have to perform an analytical
continuation of the Fourier transform
\begin{eqnarray}
          \tilde V_0(\lambda ,y) = 2\int_0^1 dx 
                                       \left[\theta (y-x) F(x,y)+
       \left\{{x\to \overline{x} \atop y \to\overline{y}}\right\}\right]_+
                                       e^{i\lambda (2x-1)},
\end{eqnarray}
which may be written as
\begin{eqnarray}
                 \tilde V_0(\lambda ,y) &=& 2\int_0^y dx
               F(x,y) \left(e^{i\lambda (2x-1)} - e^{i\lambda (2y-1)}\right)
\nonumber\\
            &&\mbox{}+2\int_0^{\overline{y}} d\overline{x}
                       F(\overline{x},\overline{y})
                 \left(e^{i\lambda (1-2\overline{x})}-
                e^{i\lambda (1-2\overline{y})}\right).
\end{eqnarray}
Both integrals are well defined and represent an analytical function with
respect to the variables $\lambda$ and $y$. In the first integral we
introduce the variable $z=x/y$ and in the second the variable
$z=\overline{x}/\overline{y}$ so that
\begin{eqnarray}
\label{513}
            \tilde V_0(\lambda ,y)=2\int_0^1 dz
           yF(zy,y)\left(e^{i\lambda (2zy-1)} - e^{i\lambda (2y-1)}\right)+
              \left\{{y\to\overline{y}\atop\lambda\to -\lambda}\right\}.
\end{eqnarray}
Obviously, $yF(yz,y)=y-1/(1-z)$ is an entire function with respect to $y$,
and the integrand
$yF(zy,y) \left(e^{i\lambda (2zy-1)} - e^{i\lambda (2y-1)}\right)$
has no singularities for $0 \leq z \leq 1$. Then the analytical
continuation of $\tilde V_0(\lambda ,y)$ is given by
\begin{eqnarray}
      \{\tilde V_0(\lambda ,y)\}_{AC}=2\int_0^1 dz
 \{yF(zy,y)\}_{AC}\left(e^{i\lambda (2zy-1)}-e^{i\lambda (2y-1)}\right)+
\left\{{y\to\overline{y}\atop \lambda \to -\lambda}\right \},
\end{eqnarray}
where $\{yF(zy,y)\}_{AC}=y-1/(1-z)$ denotes the analytical
continuation with  respect to the variable $y$.
Obviously the inverse transformation provides
\begin{eqnarray}
  V_0^{ext}(x,y)&=&\int_0^1 dz
                 \{yF(zy,y))\}_{AC}(\delta (x-2zy)-\delta (x-y))+
               \left\{{x\to\overline{x}\atop y\to \overline{y}} \right\}
\nonumber\\
                &=&\theta (1-z)\theta (z)\hbox{sign}(y) \{F(zy,y)\}_{AC}
                     |\scriptstyle z={x\over y}\displaystyle 
\nonumber \\
            &&\mbox{}-\delta (x-y) \int_0^1 dz \{yF(zy,y)\}_{AC}+
             \left\{{x\to\overline{x} \atop y
\to\overline{y}}\right\}
\nonumber \\
                &=&\left[\theta (1-z)\theta (z)\hbox{sign}(y)
         \{F(zy,y)\}_{AC}|\scriptstyle z={x\over y}\displaystyle\right]_++
                \left\{{x\to \overline{x}\atop y\to \overline{y}}\right\}.
\end{eqnarray}
So we obtain as extended BL-kernel in one-loop approximation
\begin{eqnarray}
   V_0^{ext}(x,y)&=&\gamma_0(t=2x-1,t'=2y-1)
\nonumber\\
                               &=&\theta\left(1-{x\over y}\right)
                               \theta\left({x\over y}\right)\hbox{sign}(y)
                               {x\over y}\left(1-{1\over x-y}\right) +
         \left\{{x\to \overline{x}\atop y\to \overline{y}}\right\}.
\end{eqnarray}
Note that this result can be easily checked by a direct calculation of
$\gamma_0(t,t')$.
 
The same procedure can be performed for the two-loop contributions to the
BL-kernel. The calculations are much more complicated. For convenience,
we  consider each term separately.
 
We start with the contributions coming from the functions $v_N\ ,\ v_G$
and $v_F$. They have - as in the foregoing case - the property that the
functions $yv_N(zy,y),\ yv_G(zy,y)$ and $yv_F(zy,y)$ are entire analytical
functions  of  the  variable $y$. Therefore, the extension of the
corresponding part of the BL-kernel can be performed in the same manner:
The extension of these contributions is obtained by  substituting
\begin{eqnarray}
\label{517}
\theta (y-x) \to \theta\left(1-{x\over y}\right)
                 \theta\left({x\over y}\right) \mbox{sign}(y)
\end{eqnarray}
and performing the analytical continuation of the functions $v_N,\ v_G$ and
$v_F$, e\@.g\@.
\begin{eqnarray}
\label{518} 
      v_F(x,y)=\{v_F(zy,y)\}_{AC}|\scriptstyle z={x\over y}\displaystyle.
\end{eqnarray}
 
More problems occur for the contributions
${x\over 2\overline{y}} k({x\over y})$ and ${x\over 2\overline{y}} k(x)$.
These functions  have a pole for $y = 1$. However, the Fourier
transformation of
\begin{eqnarray}
    K(x,y) = \theta (y-x) {x\over 2\overline{y}} k\left({x\over y}\right)-
                          {x\over 2\overline{y}} k(x)+
                   \left\{{x\to\overline{x}\atop y\to\overline{y}}\right\}
\end{eqnarray}
can be written in the following manner
\begin{eqnarray}
\label{519b}
\tilde K(\lambda ,y)=\int_0^1 dz {z\over\overline{y}} [y^2k(z)
            \left(e^{i\lambda (2zy-1)}-e^{i\lambda (2y-1)}\right)-\{y=1\}]+
            \left\{{y\to\overline{y}\atop\lambda\to-\lambda}\right\}.
\end{eqnarray}
Obviously, the integrand in Eq. (\ref{519b}) is an entire function, and we
can use the procedure described above.
 
The extension of $G(x,y)$ is more complicated. If we write down its Fourier
transform in the same manner as in Eq. (\ref{513}) then the integrand seems
to contain poles and cuts. These are  caused  by  the
$\overline{F}$-function and by the ln- and Li$_2$-functions, respectively.
However,  with  the help  of  identities  for  the Li$_2$-functions it is
possible to find such a representation
in the region $0\leq y\leq  1$  that  the  integrand  is  an  analytic
function (without poles and cuts) for $0\leq\mbox{Re}\ y$.
Similarly, it is possible to find a representation
where the integrand is an
analytical function for $\mbox{Re}\ y\leq 1$.
  With  other  words,  the  Fourier
transform of $G(x,y)$ is in fact an entire function, but, opposite to the
cases treated above, there does  not  exist a representation for the
integrand in  terms  of  Li$_2$-functions which is well suited for all values
of $y$. Therefore, formally we can    obtain
the  extension  by  performing the  substitution  (\ref{517}) (of course,
$\theta (y-\overline{x})\to\theta (1-{\overline{x}\over y}) \theta
({\overline{x}\over y}) \hbox{sign}(y)$) and the analytical continuation of
the  corresponding  functions,
\begin{eqnarray}
\label{520} 
G^{ext}(x,y)&=&2\theta\left(1-{x\over y}\right)\theta\left({x\over y}\right)
 \mbox{sign}(y) \{\overline{F} \hbox{Li}_2(\overline{x})+
          \overline{F}\ln(y)\ln(\overline{x})-
          F\hbox{Li}_2(\overline{y})\}_{AC}
\nonumber\\
&&\mbox +\theta\left(1-{\overline{x}\over y}\right)\theta\left({\overline{x}
          \over y}\right) \hbox{sign}(y) \Bigg\{ 2(F-\overline{F})
          \hbox{Li}_2\left(1-{x\over y}\right)+(F-\overline{F}) \ln^2 (y)
\nonumber \\
         &&\hspace{1cm}\mbox- 2F\ln(y)\ln(x) +2F\hbox{Li}_2(\overline{y})-
         2F\hbox{Li}_2(x)\Bigg\}_{AC}
         +\left\{{x\to\overline{x}\atop y\to\overline{y}}\right\}.
\end{eqnarray}
As result, we find for $y<0$ and $y>1$
terms which contain an  imaginary  part.  However, these cancel each other
exactly so that in the physically interesting regions the extension
provides  in  fact  a real function.
 
As  final  result  the extended BL-kernel reads:
\begin{eqnarray}
[\gamma (2x-1,2y-1)]_+&=&V^{ext}(x,y)
\nonumber\\
                      &=&{\alpha_s\over 2\pi}c_F[V_0^{ext}(x,y)]_++
                         \left({\alpha_s \over 2\pi}\right)^2
   c_F[c_FV_F^{ext}(x,y)+(c_G/2)V_G^{ext}(x,y)
\nonumber\\
&&\hspace{1cm}\mbox{ }+(N_F/2)V_N^{ext}(x,y)]_++O(\alpha_s^3),
\end{eqnarray}
where the extension of $V_0,\ V_F\ V_G$ and $V_N$  (\ref{57a}-\ref{Gxy})
follows  from  the  discussed  procedure  which  is  given  by  (\ref{517}), 
(\ref{518}) and (\ref{520}). So, up to the changed $\theta$-structures the 
analytic  expressions are formally identical to the originally calculated
ones.

\subsection{Consistency between the QCD Calculation of the BL- and
AP-Kernels}

An interesting application of our formula (\ref{APK}) would be a check of the
mutual consistency of the very complicated two-loop calculations of the
BL-kernel \cite{2D,2B} and the AP-kernel \cite{2A}. As it has been shown
the extended BL-kernel should contain the AP-kernel as limiting case.
Therefore,  independent calculations of both kernels are not necessary.
Otherwise,  if  they  exist,  then  they  must be consistent with each
other.

The existence of a connection between the AP- and the BL-kernels follows
already from a very simple argument. Indeed,  looking at the local
anomalous dimensions it seems to be obvious that the general anomalous
dimension matrix $V_{nm}=\gamma_{nm}-2\gamma_\psi\delta_{nm}$ corresponding
to the BL-kernel \cite{MM}
\begin{eqnarray}
\int_0^1 dx x^nV_{BL}(x,y) = \sum_{m=0}^n V_{nm}y^m
\end{eqnarray}
contains the diagonal matrix elements corresponding to the AP-kernel,
\begin{eqnarray}
\int_{-1}^1 dz  z^n P(z)=V_{nn}, \quad P(z) = {1\over 2\pi i}
\int_{-c-i \infty}^{-c+i \infty} dn  z^{-n-1} V_{nn}.
\end{eqnarray}
If the local anomalous dimensions would determine the corresponding kernels
then it should be possible to determine the AP-kernel from the
given BL-kernel. Starting from this observation there were several
incomplete trials to solve this problem \cite{2D}.
 
Here, we shall determine the
AP-kernel from the known extended BL-kernel. In the language of the local
anomalous dimension we can say that in this way an independent check for the
correctness of the diagonal anomalous dimension will be performed. A check
for the nondiagonal elements of the anomalous dimension matrix has been
carried out by exploiting consequently conformal symmetry breaking \cite{DM}.
Therefore, the considerations given in the following enable us to conclude
that  the two-loop approximation of the BL-kernel has been correctly
computed.

To obtain the AP-kernel practically, we have to apply Eq. (\ref{APK}) which we
reproduce here for convenience
\begin{eqnarray}
\label{524}
           P(z)=\lim_{\tau\to 0}|\tau |^{-1} [\gamma (2x-1,2y-1)]_+
|\scriptstyle x={z\over\tau},y={1\over\tau}\displaystyle
\end{eqnarray}
Let us remark  that according to this equation it is just the added new
region ($|t|,|t'|>1$) which is essential for the determination of the
AP-kernel.

For technical reasons, we  first discuss the $\theta$-structure and then
we consider the
limiting procedure for the coefficient functions contained in Eqs. 
(\ref{58a}-\ref{Gxy}).
The limits $\tau \to 0$ with $x=z/\tau $, $y=1/\tau $ for the
$\theta$-functions are
\begin{eqnarray}\left. {\theta ({x\over y})\theta (1-{x\over y})\atop
\theta ({1-x\over 1-y})\theta (1-{1-x\over 1-y})}\right\}
\quad &\to& \theta (z)\theta (1-z), \\
\left. {\theta ({1-x\over y})\theta (1-{1-x\over y})\atop
\theta ({x\over 1-y})\theta (1-{x\over 1-y})}\right\}
\quad &\to& \theta (-z)\theta (1+z).
\end{eqnarray}
It is important for further calculations that "related $\theta$-structures",
i.e. structures which turn into each other under
$x\leftrightarrow\overline{x}=1-x$ and
$y\leftrightarrow\overline{y}=1-y$ - have the same limit.
We note, that all expressions have the following typical structures
\begin{eqnarray}
S_1(x,y)=\theta\left({x\over y}\right)\theta\left(1-{x\over y}\right)
                      \mbox{sign}(y) T_1(x,y)
             +\theta\left({\overline{x}\over \overline{y}}\right)
              \theta\left(1-{\overline{x}\over \overline{y}}\right)
              \mbox{sign}(\overline{y})T_1(\overline{x},\overline{y})
\nonumber\\
\end{eqnarray}
or
\begin{eqnarray}
             S_2(x,y)= \theta\left({\overline{x}\over y}\right)
                        \theta\left(1-{\overline{x}\over y}\right)
                        \mbox{sign}(y) T_2(x,y)
                       +\theta\left({x\over\overline{y}}\right)
                        \theta\left(1-{x\over\overline{y}}\right)
                        \mbox{sign}(\overline{y})
                        T_2(\overline{x},\overline{y}).
\nonumber\\
\end{eqnarray}
So we have to determine the following limits:
\begin{eqnarray}
    \lim_{\tau\to 0}|\tau |^{-1} S_1\left({z\over\tau},{1\over\tau}\right) =
              \theta (z)\theta (1-z) \lim_{\tau\to 0} \tau^{-1}
              \left\{T_1\left({z\over\tau},{1\over\tau}\right)
              -T_1\left(1-{z\over\tau},1-{1\over \tau }\right)\right\}
\nonumber\\
\end{eqnarray}
and
\begin{eqnarray}
     \lim_{\tau\to 0}|\tau |^{-1}S_2\left({z\over\tau},{1\over\tau}\right) =
              \theta (-z)\theta (1+z) \lim_{\tau\to 0}\tau^{-1}
              \left\{T_2\left({z\over\tau},{1\over\tau}\right)
               -T_2\left(1-{z\over\tau},1-{1\over\tau}\right)\right\}.
\nonumber\\
\end{eqnarray}
As short notation we introduce the symbol LIM,
\begin{eqnarray}
         \mbox{LIM} T(x,y)=\lim_{\tau\to 0}\tau^{-1}
                       \left\{T\left({z\over\tau},{1\over\tau}\right)-
                       T\left(1-{z\over\tau},1-{1\over\tau}\right)\right\}.
\end{eqnarray}
The determination of these limits will be performed in  Appendix E.

Putting  together  all results obtained there we get an expression for
the  AP-kernel  following from the extended BL-kernel 
(\ref{56}-\ref{Gxy}). In analogy with Eq. (\ref{56}) we write
\begin{eqnarray}
\label{532}
P(z)&=&{\alpha_s\over 2\pi}c_F[P_0(z)]_++\left({\alpha_s\over 2\pi}\right)^2
         c_F [c_F P_F(z)+ (c_G/2)P_G(z)+(N_F/2)P_N(z)]_+
\nonumber\\
& &\mbox{\ }+O(\alpha_s^3).
\end{eqnarray}
Let us compute each term separately
\begin{eqnarray}
                P_0(z)&=&\lim_{\tau\to 0}|\tau|^{-1}
              V_0\left({z\over\tau},{1\over\tau}\right)
\nonumber\\
          &=&\theta (z)\theta (1-z) {1+z^2\over 1-z} 
\end{eqnarray}
according to (\ref{E1b}).
\begin{eqnarray}
          P_N(z)&=&\lim_{\tau\to 0}|\tau |^{-1}
                   V_N\left({z\over\tau},{1\over\tau}\right)
\nonumber\\
                &=&-\theta (z)\theta (1-z){2\over 3}\left[{1+z^2\over 1-z}
                    \left(\ln(z)+{5\over3}\right)+2(1-z)\right]
\end{eqnarray}
because of (\ref{E1a}-\ref{E1c}).
\begin{eqnarray}
          P_G (z)&=&\lim_{\tau\to 0}|\tau |^{-1}
                    V_G\left({z\over\tau},{1\over\tau}\right)
\nonumber\\
         &=&\theta (z)\theta (1-z)\Bigg[{1+z^2\over 1-z}\left(\ln^2 (z)+
                 {11\over 3}\ln(z)+{67\over 9}-{\pi^2 \over 3}\right)
\nonumber\\
       &&\qquad\qquad\qquad\qquad\mbox{}+2(1+z)\ln(z)+{40\over 3}(1-z)\Bigg]
\nonumber\\
            &&\mbox{}+\theta (-z)\theta (1+z)\Bigg[2{1+|z|^2\over 1+|z|}
                    \Bigg(\hbox{Li}_2\left({|z|\over 1+|z|}\right)-
              \hbox{Li}_2\left({1\over 1+|z|}\right)+{1\over 2}\ln^2(|z|)
\nonumber\\
                 &&\qquad\qquad\qquad\qquad\mbox{}-\ln(|z|)\ln(1+|z|)\Bigg)
                   +2(1+|z|)\ln(|z|)+4(1-|z|)\Bigg].
\nonumber\\
\end{eqnarray}
Here we have taken into account (\ref{E1a}-\ref{E1c},\ref{E1g},\ref{E1h}).
For the last term we obtain the result:
\begin{eqnarray}
\label{536}
          P_F(z)&=&\lim_{\tau\to 0} |\tau |^{-1}
                   V_F\left({z\over\tau},{1\over\tau}\right) 
\nonumber\\
                &=&\theta (z)\theta (1-z)\left[-{1+z^2\over
1-z}\left(\ln^2(z)+
                    {3\over 2}\ln(z)+2\ln(z)\ln(1-z)\right)
                    \right.
\nonumber\\
                 &&\hspace{1cm}\left. +{1+3z^2\over 2(1-z)}\ln^2(z) 
                     -{1\over 2}(3+7z)\ln(z)
                   -5(1-z)\right] \nonumber \\
              &&\mbox{}-\theta (-z)\theta (1+z)\Bigg[2{1+|z|^2\over 1+|z|}
                      \Bigg(\hbox{Li}_2\left({|z|\over 1+|z|}\right)-
              \hbox{Li}_2\left({1\over 1+|z|}\right)+{1\over 2}\ln^2(|z|)
        \nonumber\\
       &&\qquad\qquad\qquad\qquad\mbox{}-\ln(|z|)\ln(1+|z|)\Bigg)
                       +2(1+|z|)\ln(|z|)+4(1-|z|)\Bigg].
\nonumber\\
\end{eqnarray}
The obtained result for the AP-kernel (\ref{532}-\ref{536}) is completely
equivalent to
that of Ref. \cite{2A}. It also includes in a natural way the second order
contributions stemming from the internal anti-quark lines.
 
Let us add a remark concerning the regularization prescription of the
considered kernel. It is very important, that our regularization
prescription for the extended BL-kernel turns  after performing the
limiting process, automatically into the well accepted regularization
prescription for the AP- kernel. If we start with (\ref{RP}) and (\ref{524})
\begin{eqnarray}
\label{537}
                \lim_{\tau\to 0}|\tau |^{-1}
                \left[\gamma\left({z\over\tau},{1\over\tau}\right)-
                \delta\left({z\over\tau}-{1\over\tau}\right)
                \int_{-\infty}^{\infty} dx'
                \gamma\left(2x'-1,{2\over \tau}-1\right)\right]
\end{eqnarray}
then the limit of the $\theta$-structure restricts the integration  range  in
Eq. (\ref{537}). The substitution $x' = z'/\tau$ leads to
\begin{eqnarray}
            \lim_{\tau\to 0}|\tau |^{-1}
                 \left[\gamma\left({z\over\tau},{z\over\tau}\right)
                 -\delta (z-1) \int_{-\infty}^{\infty} dz'
                \gamma\left({z'\over\tau},{1\over\tau}\right)\right]=
                 P(z)-\delta (z-1) \int    dz' P(z')
\nonumber\\
\end{eqnarray}
This completes our considerations. 
 
\pagebreak[4]
\appendix

\section{Virtual Compton  Scattering Amplitude \\
in Leading Approximation}

The  aim  of  this Appendix is to study the behaviour of the
helicity  amplitudes  of  the  Compton  scattering  amplitude  in  the
generalized Bjorken region. The helicity amplitudes are defined by
\begin{eqnarray}
                 T(\lambda ',\lambda)=
  \varepsilon_2^\mu(\lambda ') T_{\mu\nu} \varepsilon_1^\nu (\lambda),
\end{eqnarray}
where $\varepsilon_i(\lambda )$ denotes the polarization vector of the
virtual photon $(i=1,2)$:
\begin{eqnarray}
                 \varepsilon_i^\mu(\lambda )&=&
                                            (0,\vec\varepsilon_i(\lambda )),
            \quad\vec\varepsilon_i(\lambda )\vec q_i=0,
            \quad\vec\varepsilon_i(\lambda )\vec\varepsilon_i(\lambda ') =
                                               \delta_{\lambda \lambda '},
\quad\hbox{for}\quad \lambda ,\lambda '= 1,2,
\nonumber\\
                 \varepsilon_i^\mu(\lambda =3) &=&
                         |q_i^2|^{-1/2}(|\vec q_i|,q_i^0\vec q_i/|\vec q_i|),
\nonumber\\
                               &=&|q_i^2|^{-1/2}(q_i^\mu-q_i^2c_i^\mu/c_iq_i),
           \quad c_i^\mu = (1,-\vec q_i/|\vec q_i|),
\nonumber\\
               \varepsilon_i^\mu (\lambda =0) &=&
                                             |q_i^2|^{-1/2} q_i^\mu .
\end{eqnarray}
It will be  shown that in leading order they can be expressed as
\begin{eqnarray}
\label{AzBR}
   T(\lambda ',\lambda)\approx \cases{
{1\over 2}\varepsilon_2^\mu (\lambda ')
 \varepsilon_{1\mu}(\lambda) T_\nu^{\mbox{ }\nu}
        &for $\lambda ',\lambda =1,2,$ \cr
0 &otherwise, \cr}
\end{eqnarray}
so that in section III we need to consider the more simple
expression $T_\nu^{\mbox{ }\nu}$ only.

To get these results at first we  remark that because of the
current conservation $q_2^\mu T_{\mu\nu} =0$ and $T_{\mu\nu} q_1^\nu =0$
the helicity amplitudes
$T(\lambda ',\lambda )$ vanish for  $\lambda '=0$ or $\lambda =0$. For the
same reason a kinematical suppression of the longitudinal
degrees of freedom occurs, for example
\begin{eqnarray}
     \varepsilon_2^\mu (\lambda =3) T_{\mu\nu}=
                             |q_2^2|^{-1/2} q_2^\mu T_{\mu\nu}
                     \pm {\sqrt{|q_2^2|}\over c_2q_2}c_2^\mu T_{\mu \nu} =
                     \pm {\sqrt{|q_2^2|}\over c_2q_2}c_2^\mu T_{\mu \nu}
\quad\mbox{for } q^{2\ <}_{2\ >}\ 0
\end{eqnarray}
is suppressed because the factor
\begin{eqnarray}
                {\sqrt{|q_i^2|}\over c_iq_i}=
    {\sqrt{|(q\pm P_- /2)^2|}\over c_i(q\pm P_- /2)}\approx
    {\sqrt{Q^2}\over c_iq}=O\left({1\over\sqrt{Q^2}}\right)
     \quad i=1,2.
\end{eqnarray}
vanishes  asymptotically.

For  further  simplification  we  need  explicite representations and
approximations of the scattering amplitude. Because of our kinematical
assumptions,  in  $x$-space  the  scattering  amplitude (\ref{ScA}) is
dominated  by  contributions  arising  from  the  neighbourhood of the
light-cone. So it is possible to apply the light-cone expansion of the
current  product.  Instead of the standard expansion in terms of local
operators  we  use  the  more  effective  expansion based on light-ray
operators  \cite{AZ,Za}.  This  expansion   relies on a perturbative
expansion  of  the  product  of the electromagnetic currents, too. The
crucial  difference  between both expansions is the following: Whereas
usually  the  light-cone  expansion  follows  from  a  suitable Taylor
expansion   leading  to  the  "local"  light-cone  operators,  here  a
corresponding  Fourier  transform generates the nonlocal ("light-ray")
operators. In leading order we obtain \cite{Ge}
\begin{eqnarray}
\label{AOPE}
TJ_{\mu}\left({x \over 2}\right)J_{\nu}\left(- {x \over 2}\right)
          &\approx& \int     d\kappa_+d\kappa_-
                     F^a(x^2,\kappa_+,\kappa_-,\mu^2)
          \Gamma_{\mu\nu}^{\mbox{ } \mbox{ }\alpha\beta}\tilde x_\alpha
   O_\beta^a(\kappa_+\tilde x,\kappa_-\tilde x )_{(\mu^2)}
\nonumber\\
            &&\mbox{}+\int     d\kappa_+ d\kappa_-
          \hat F^a(x^2,\kappa_+,\kappa_-,\mu^2)
          \epsilon_{\mu\nu}^{\mbox{ \ }\alpha\beta}\tilde x_\alpha
\hat O_\beta^a(\kappa_+\tilde x,\kappa_-\tilde x)_{(\mu^2)},
\nonumber\\
\end{eqnarray}
where
$\Gamma_{\mu\nu}^{\mbox{\ \ }\alpha\beta}=g_{\mu\nu}g^{\alpha\beta} -
g_\mu^{\mbox{ }\alpha}g_\nu^{\mbox{ }\beta}-
g_\mu^{\mbox{ }\beta}g_\nu^{\mbox{ }\alpha}$ and 
$\epsilon_{\mu\nu}^{\mbox{\ \ }\alpha\beta}$ denotes the $\epsilon$-tensor
and
\begin{eqnarray*}
\tilde x(x,\rho )= x +
\rho {x\rho \over \rho^2}\left(\sqrt{1-{x^2 \rho^2\over (x\rho )^2}}-1\right)
\end{eqnarray*}
is a light-like vector determined by $x$ and parameterized by a fixed
constant vector $\rho$.
In the tree approximation we obtain for the coefficient functions
$F^a (x^2,\kappa_+,\kappa_-,\mu^2)$ and
$\hat F^a(x^2,\kappa_+,\kappa_-,\mu^2)$:
\begin{eqnarray}
\label{AKF}
              F^a(x^2,\kappa_+,\kappa_-)&=&
              {ic^a\over 2\pi^2(x^2-i\epsilon )^2} \delta (\kappa_+)
            \left(\delta (\kappa_--1/2)-
                  \delta (\kappa_-+1/2)\right) 
\nonumber\\
              \hat F^a(x^2,\kappa_+,\kappa_-)&=&
               {ic^a\over 2\pi^2(x^2-i\epsilon)^2} \delta (\kappa_+)
              \left(\delta (\kappa_--1/2)+\delta (\kappa_- +1/2 )\right),
\nonumber \\
       c_a &=& {2\over 9} \delta_{a0}+{1\over 6}\delta_{a3}+
                                            {1\over 6\sqrt{3}}\delta_{a8}
               \quad \hbox{for flavour SU(3)}.
\end{eqnarray}
In this approximation the gluon operators are automatically suppressed,
and only the quark operators contribute:
\begin{eqnarray}
          O^a_\beta(\kappa_+\tilde x,\kappa_-\tilde x) &=&
  \;:\!\overline{\psi }(\kappa_1 \tilde x) \gamma_\beta\lambda^a
      U(\kappa_1 \tilde  x,\kappa_2 \tilde x) \psi(\kappa_2 \tilde  x)\!:,\\
     \hat O^a_\beta (\kappa_+\tilde x,\kappa_-\tilde x) &=&
  \;:\!\overline{\psi }(\kappa_1 \tilde x) i\gamma^5 \gamma_\beta\lambda^a
      U(\kappa_1 \tilde x,\kappa_2\tilde x)\psi(\kappa_2 \tilde x)\!:,
                           \quad\kappa_\pm =(\kappa_2\pm \kappa_1)/2.
\end{eqnarray}
Here,  $U(\kappa_1\tilde  x,\kappa_2\tilde  x)$  is the standard phase
factor  (\ref{DO}),  and  $\lambda^a$  determines the flavour content,
additionally  we  defined  $\lambda^0  \equiv 1$ so that the summation
over  the index $a$ includes also the flavour singlet case. The choice
$\mu^2=Q^2$  for  the renormalization point justifies the perturbative
calculation of the coefficient functions.

The  nonperturbative  part  of  the  Compton  scattering  amplitude is
contained in the matrix elements of the quark operators. Leaving aside
the  nonessential  $\kappa_+$-dependent  exponential [in fact it drops
out because of $\delta (\kappa_+)$ in Eqs. (\ref{AKF})]
\begin{eqnarray*}
       <P_2|O^a_\beta(\kappa_+ \tilde x,\kappa_-\tilde x)|P_1>=
                           <P_2|O^a_\beta(\kappa_-\tilde x)|P_1>
                                           e^{i\kappa_+ (\tilde x P_-)}
\end{eqnarray*}
we write down the general structure of the matrix elements of these
operators between scalar or pseudoscalar mesons or spin averaged baryon
states
\begin{eqnarray}
\label{AM1}
          <P_2|TO^a_\beta(\kappa_-\tilde x)|P_1> &=&
                            P_{+\beta} t_1^a (\kappa_- \tilde x,P_i,\mu^2)+
                            P_{-\beta} t_2^a(\kappa_-\tilde x,P_i,\mu^2)
\nonumber\\
         &&\mbox{}+\kappa_-\tilde x_\beta M^2t^a_{nl}(\kappa_-\tilde
x,P_i,\mu^2),
\\
\label{AM2}
         <P_2|T\hat O^a_\beta(\kappa_-\tilde x)|P_1>&=&
                  \epsilon_{\beta \gamma \delta \epsilon}
             P_+^\gamma P_-^\delta \kappa_- \tilde x^\epsilon
                  \hat t^a(\kappa_-\tilde x,P_i ,\mu^2 ).
\end{eqnarray}
The mass factor $M^2=P_i^2$ in the third term of Eq. (\ref{AM1}) has to
be  introduced  for dimensional reasons. This term does not contribute
in   leading   order.  Inserting  Eqs. (\ref{AOPE}), (\ref{AM1}), (\ref{AM2})  into
Eq.  (\ref{ScA}) we are able to write formally
\begin{eqnarray}
\label{AZR}
                     T_{\mu\nu}(P_+,P_-,q)&=&
   \Gamma_{\mu\nu\alpha\beta} \left(P_+^\beta{\partial\over\partial q_\alpha}
        T_1 (P_+,P_-,q)+ P_-^\beta{\partial\over\partial q_\alpha}
                            T_2(P_+,P_-,q)\right)
\nonumber\\
            &&\mbox{}-i\epsilon_{\mu\nu\alpha\beta}
    \epsilon^\beta_{\mbox{\ } \gamma\delta\epsilon} P_+^\gamma P_-^\delta
       {\partial\over\partial q_\alpha}{\partial\over\partial q_\epsilon}
                            \hat T(P_+,P_-,q),
\end{eqnarray}
where
\begin{eqnarray}
           T_i(P_+,P_-,q)= \int d^4x\int  d\kappa_+ d\kappa_-
                   e^{ixq+ i\kappa_+(\tilde xP_-)}
        F^a(x^2,\kappa_+,\kappa_-) t^a_i(\kappa_-\tilde x,P_i,Q^2).
\end{eqnarray}
A corresponding formula is valid for $\hat T(P_+,P_-,q)$.
The r.h.s. of Eq. (\ref{AZR}) can be simplified further: Since
\begin{eqnarray}
      {\partial\over\partial q_\alpha} T_i(P_+,P_- ,q) &=&
         \left(2q^\alpha {\partial\over\partial q^2} +
\left({\partial\over\partial q_\alpha}\xi\right){\partial\over\partial\xi}
+\left({\partial\over\partial q_\alpha}\eta\right){\partial\over\partial\eta}
                 \right) T_i(\xi ,\eta ,Q^2)
\nonumber\\
     &=&\left(-2q^\alpha{\partial\over\partial Q^2}-
         {2q^\alpha +\xi P_+^\alpha\over qP_+} {\partial\over\partial\xi}+
         {P_-^\alpha -\eta P_+^\alpha \over qP_+}{\partial\over\partial\eta}
                 \right)T_i(\xi ,\eta ,Q^2)
\nonumber\\
\end{eqnarray}
we can estimate
\begin{eqnarray}
              {\partial\over\partial q_\alpha} T_i(P_+,P_- ,q)
               &\approx & -2q^\alpha\left({\partial\over\partial Q^2}+
                        {1\over qP_+}{\partial\over\partial\xi}
                 \right) T_i(\xi ,\eta ,Q^2) 
\nonumber\\
&\approx & -{2q^\alpha\over Q^2} \left(Q^2 {\partial\over\partial Q^2}+
                        \xi {\partial\over\partial\xi}
                 \right) T_i(\xi ,\eta ,Q^2)
                 \qquad\mbox{for}\quad Q^2,qP_+,qP_-\to\infty.
\nonumber\\
\end{eqnarray}
Using
\begin{eqnarray}
       \varepsilon_i(\lambda )q=-\varepsilon_i(\lambda )(q_i+(-1)^iP_- )
                            =-(-1)^i\varepsilon_i(\lambda )P_-
      \quad \mbox{for}\quad\lambda = 1, 2,
\end{eqnarray}
and the explicit expression for $\Gamma_{\mu\nu\alpha\beta} $,
we obtain after some algebra the leading contributions for the transversal
helicity amplitudes:
\begin{eqnarray}
\label{AHA}
      T(\lambda ',\lambda)\approx
              \varepsilon_2^\mu (\lambda ')\varepsilon_{1\mu} (\lambda)
                       \left(P_+{\partial\over\partial q}T_1
                        +P_-{\partial\over\partial q} T_2 \right).
\end{eqnarray}
Here we  took  into  consideration,  that  the  $\hat  T$  amplitude  gives
asymptotically  nonleading  contribution,  which   follows   from   similar
arguments. Using Eq. (\ref{AZR}), a direct calculation of the trace
$T_{\mu\nu}$ shows
\begin{eqnarray}
\label{ATA}
        T_\mu^{\mbox{\ } \mu}&\approx&
              2\left(P_+{\partial\over\partial q}T_1
             +P_-{\partial \over \partial q} T_2 \right).
\end{eqnarray}
Comparing Eqs. (\ref{AHA}) and (\ref{ATA}), we obtain the desired result
(\ref{AzBR}).

\section{Properties of Anomalous Dimensions \\
of Light-Ray Operators}
 
This  Appendix  is  devoted  to  a  general  discussion of the support
properties   of   the  anomalous  dimensions  of  light-ray  operators
(\ref{DOV}), 
\begin{eqnarray}
\label{B1}
O^a(\kappa_+ ,\kappa_- ;\tilde n) =  :\! \overline{\psi }(\kappa_1 \tilde
n) (\tilde n \gamma) \lambda^a U(\kappa_1 \tilde n,\kappa_2 \tilde n)
\psi(\kappa_2 \tilde n)\!:, \quad \kappa_\pm = (\kappa_2 \pm \kappa_1 )/2,
\quad {\tilde n}^2 =0.
\nonumber\\
\end{eqnarray}
These operators differ from the original ones (\ref{DO}) by a translation.
Their renormalization group equation reads \cite{Br}
\begin{eqnarray}
\label{BRGE}
\mu {d \over d\mu}O^a (\underline{\kappa };\tilde n)_{(\mu^2)}  = \int d^2
\underline{\kappa'} \left( \gamma (\underline{\kappa },
\underline{\kappa' };g(\mu^2 ) ) - 2\gamma_\psi( g(\mu^2 )) \delta^{(2)}
(\underline{\kappa }- \underline{\kappa' }) \right) O^a
(\underline{\kappa' };\tilde n)_{(\mu^2)}.
\nonumber\\
\end{eqnarray}
For convenience we divide the anomalous dimension of the operator 
into the anomalous dimension of the 1PI vertex function 
(with two external momenta)  
$\gamma (\underline{\kappa } ,\underline{\kappa ' }) =  \gamma
(\kappa_+,\kappa_-,\kappa '_+ ,\kappa '_- ) $ 
and a part which is proportional to the anomalous dimension of
the quark field $ \gamma_\psi $ . As a shorthand notation we use
$O^a(\underline{\kappa};\tilde n)=O^a(\kappa_+,\kappa_-;\tilde n)$, 
$d^2 \underline{\kappa ' } = d\kappa '_+ d\kappa '_- $ and  $\delta^{(2)}
(\underline{\kappa  }-\underline{\kappa '  }) = \delta (\kappa _+ -
\kappa '_+)\linebreak[2] \delta (\kappa '_- - \kappa '_-)$.

In the following, we  investigate the anomalous dimensions $\gamma
(\underline{\kappa } ,\underline{\kappa ' } )$ 
of the above defined operator 
using the $\alpha$-representation for 1-particle-irreducible
(1PI) diagrams containing
this operator insertion.
As  result,  we shall  obtain  the  support
restriction (\ref{BSR}), i.\ e.  $|w_\pm |\leq 1, |w_+\pm w_-|\leq 1$ with
$w_+=(\kappa_+'-\kappa_+ )/ \kappa_-,\ w_- =\kappa_-' /\kappa_-$ and the
special variable dependence $\gamma (\underline{\kappa},\underline{\kappa
'})=\kappa_-^{-2}\gamma (w_+,w_-)$. Additionally, from the 
transformation properties of
the operators (\ref{B30}) under charge conjugation it follows
$\gamma (w_+,w_-) = \gamma (-w_+,w_-)$. 
In all what follows we give the
more technical proof of these results.
 
In principle, we have to investigate the renormalization of gauge invariant
operators on the light-cone. Thereby we should have in mind:
\begin{enumerate}
\item  It  is essential that these nonlocal operators are defined on a
   light-ray.   This  changes  their  renormalization  properties  and
   induces  a  close  correspondence to the local operators known from
   the standard light-cone expansion \cite{RO}.
\item From the renormalization procedure of gauge
invariant operators it is well-known that during the renormalization
 process there may appear gauge variant and ghost operators. In the
   complete renormalization matrix they appear in a triangular form, so
   that the anomalous dimensions of the physical operators remain
   unchanged. Moreover these operators have vanishing matrix elements
   between physical particle states. For this reason, we do not need to
   consider such contributions \cite{giO}.
\item Up to these restrictions
the operators (\ref{B1}) form a complete operator basis of minimal twist and
   minimal dimension.
\end{enumerate}

The anomalous dimension can be determined from the 1PI vertex function of
the  operator to be studied \cite{Za,Bo}. For instance,
by using dimensional regularization it can be shown that
in the minimal subtraction scheme the following simple
rule is valid \cite{D1}
\begin{eqnarray}
\label{B3}
\gamma (\underline{\kappa } ,\underline{\kappa '};g)=g {\partial \over
\partial g}Z^{[1]}(\underline{\kappa } ,\underline{\kappa '} ;g)
\end{eqnarray}
where $Z^{[1]}$ can be obtained from the residue of the
1PI vertex function
$R'G(p_1, p_2 ;g;\epsilon |O^a (\underline{\kappa }))$ with
respect to the parameter $\epsilon = (4-n)/2$.
In detail, by
$R' G(p_1, p_2 ;g;\epsilon |O^a (\underline{\kappa
}))$ we denote  the 1PI vertex function (with two external momenta $p_1$ and
$p_2$) containing the operator insertion where all subdiagrams are
 renormalized, and only the overall renormalization is not carried out. The
$\epsilon$-expansion of this function reads \cite{D2}
\begin{eqnarray}
\label{B4}
R' G(p_1, p_2 ;\epsilon |O^a (\underline{\kappa };\tilde n)) = G^{[0]}(p_1,
p_2|O^a (\underline{\kappa };\tilde n))  +  {1  \over  \epsilon} G^{[1]}(p_1,
p_2|O^a (\underline{\kappa };\tilde n)) + \cdots .
\end{eqnarray}
In general, the $1/\epsilon$ term is given as an integral with respect to
the $\underline{\kappa}$-variables of the $Z$-factor with the vertex
\begin{eqnarray}
\label{B5}
O^a_V(\underline{\kappa};\tilde n;\underline{p})=(\tilde
n\gamma )\lambda^a e^{i\kappa_+(\tilde np_+)+i\kappa_- (\tilde np_-)}
\end{eqnarray}
of the bare operator (\ref{B1}). We have, therefore,
\begin{eqnarray}
G^{[1]}(p_1,p_2|O^a(\underline{\kappa};\tilde n))=\int d^2
\underline{\kappa '}  Z^{[1]}(\underline{\kappa } ,\underline{\kappa '})
O^a_V (\underline{\kappa' };\tilde n;\underline{p}).
\end{eqnarray}
Using the special form (\ref{B5}) of the operator vertex we can write
\begin{eqnarray}
\label{B7}
(\tilde n \gamma ) \lambda^a  Z^{[1]}(\underline{\kappa}, \underline{\kappa
'})=\int {d (\tilde np_+) \over 2 \pi}\int {d(\tilde np_-)
\over 2 \pi}  e^{-i\kappa'_+ ({\tilde  np_+})-i\kappa'_- ({\tilde
np_-})}\, G^{[1]}(p_1, p_2|O^a (\underline{\kappa};\tilde n)).
\nonumber\\
\end{eqnarray}
Because  of  Eqs. (\ref{B3})  and  (\ref{B4})  the  investigation  of the
support properties  of  the  anomalous dimension requires the knowledge of
the support properties of all possible diagrams contributing to $R' G(p_1,
p_2  ;g;\epsilon |O^a (\underline{\kappa }))$.

The standard method for
the  investigation  of  Green's  functions  is  the application of the
$\alpha$-re\-pre\-sen\-ta\-tion.  We use the following formula for the
propagator
\begin{eqnarray} {P(k) \over m^2-k^2-i\rho}=\lim_{\xi \to 0
}  P  \left({1  \over  2i}{\partial  \over  \partial  \xi  } \right)
i\int_0^\infty  d\alpha    e^{i\alpha (k^2-m^2+i\rho)+i2\xi k}\, ,
\end{eqnarray}  
where  $P(k)$  denotes  a well spezified  polynomial in the
momentum $k$. Also a
complicated matrix structure of the vertices can be included into this
polynomial. With these remarks the  $\alpha$-representation for  a 1PI
diagram with $L$ internal lines and $V$ vertices and two external legs
can be written as \cite{Za}
\begin{eqnarray}
G_{\Gamma }(p_1,p_2)=\int_0^\infty d\underline{\alpha} \prod_{l=
1}^L P_l \left({1\over 2i}{\partial \over \partial \xi_l} \right)
e^{-i\alpha_l (m_l -i\rho )} f_\Gamma (\underline{\alpha} ; \underline{p}
;\underline{\xi})|\scriptstyle \underline{\xi}=0 \displaystyle ,
\end{eqnarray}
where
\begin{eqnarray}
f_\Gamma (\underline{\alpha};\underline{p};\underline{\xi})={\mbox{const.}
\over {D^{n/2}(\underline{\alpha})}} e^{i\left\{\sum_{i,j=1}^2p_i  A_{ij}
(\underline{\alpha}) p_j + 2 \sum_{l=1}^L \sum_{i=1}^2 \xi_l B_{li}
(\underline{\alpha}) p_i - \sum_{k,l=1}^L \xi_k K_{kl} (\underline{\alpha})
\xi_l \right\}}.
\nonumber\\
\end{eqnarray}
Thereby, the regularization of the ultra-violet divergencies is ensured by
dimensional regularization, i.e. by an analytical continuation
from $n=4$ to
$n=4-2\epsilon, \epsilon > 0$. For later convenience, we have displayed
the polynomial structure with respect to
 variables $p_i$ and $\xi_l$ explicitly and
introduced coefficients depending on $\alpha$-parameters only. Thus, our
notation differs slightly from that of Ref. \cite{Za}; furthermore, in
contrast to Ref. \cite{Za} we have
 absorbed the overall factor $1/D$ of the exponent into the definition of
$A_{ij}$, $B_{li}$ and $K_{kl}$. Anyway, the coefficients $A_{ij}$,
$B_{li}$ and $K_{kl}$ may be simply deduced from the general rules given
there.

For our purpose we need diagrams containing one insertion of the light-cone
operator (\ref{B1}). Of course, the vertex corresponding to this operator is
complicated. However, if we consider QCD in axial gauge $\tilde n^\mu A_\mu
= 0$, then the phase factor $U(\kappa_1 \tilde n,\kappa_2 \tilde n)$
 does not contribute, and only the bare vertex (\ref{B5}) remains. Moreover,
ghost operators do not exist. The  drawback of this procedure is the
comparatively  complicated  structure  of  the  gluon   propagator
\cite{KS}
\begin{eqnarray}
D_{ab}^{\mu \nu } = \delta_{ab} {-i \over {k^2+i\rho}} \left[g^{\mu \nu  }
- {{\tilde n^\mu k^\nu + \tilde n^\nu k^\mu} \over 2} \left({1 \over {\tilde
nk+i\tau}}+{1 \over {\tilde nk-i\tau}}\right) \right].
\end{eqnarray}
The  $\tilde  n$-dependent  terms contained in the second part of this
propagator  need  additional  considerations. For simplicity we choose
for the regularization of the unphysical pole term $ (\tilde n k)^{-1}
$  the  principal  value  prescription.  It  is well-known that this
procedure  generates nonlocal
counter terms. This problem can be avoided by using the Mandelstam or
Leibbrandt  prescription  \cite{Ma}.  However, in our case this is not
necessary, because  the considered physical quantities are independent
of  the  gauge  fixing  procedure  and  therefore, do not depend on the
prescription of the unphysical poles. As usual, the spurios divergences
are subtracted by hand (see also the two-loop calculation in Ref. \cite{2D}).
Below we shall show  that the contributions of the unphysical poles
do not change the support property of the anomalous dimension. So we
shall ignore them; in fact this means that we study the scalar theory first.
 
The operator vertex (\ref{B5})  may be included into each diagram as a
special first vertex connected through the lines 1 and 2 to the remaining
part of the diagram. In the  $\alpha$-representation, we have
\begin{eqnarray}
G_\Gamma (\underline{p}|O(\underline{\kappa };\tilde n)) = O_V
 \left( \underline{\kappa};  \tilde  n;  {1  \over  2i}\underline{
{\partial   \over \partial \xi}}\right) G_\Gamma  (\underline{p},
\underline{\xi})|\scriptstyle \underline{\xi}=0 \displaystyle ,
\end{eqnarray}
with
\begin{eqnarray*}
O_V  (\underline{\kappa};\tilde n;\underline{p}) = e^{i\kappa_+(\tilde
np_+) +i\kappa_- (\tilde np_-)}.
\end{eqnarray*}
The differential operator in the exponential shifts the variables
according to $\xi_1 \rightarrow \xi_1+{\kappa_1\over 2} \tilde n$ and $\xi_2
\rightarrow \xi_2 + {\kappa_2 \over 2} \tilde n $. This shift can be taken
into account explicitly, so that the $\alpha$-representation with the
special operator insertion takes the form
\begin{eqnarray}
      G_\Gamma (\underline{p}|O(\underline{\kappa };\tilde n))&=&
                               \int_0^\infty d\underline{\alpha}
    \prod_{l=1}^L P_l \left({1\over 2i}{\partial \over\partial\xi_l}\right)
                              e^{-i\alpha_l (m_l - i\rho)}
    f_\Gamma (\underline{\alpha};\underline{p};\underline{\xi }) 
\nonumber\\
    &&\times \exp i\left\{\sum_{i,j=1}^2 \kappa_i B_{ij}(\tilde np_j)-
     \sum_{i=1}^2 \sum_{l=1}^L \kappa_i K_{il} (\tilde n\xi_l)\right\}
                         |\scriptstyle \underline{\xi}=0 \displaystyle .
\end{eqnarray}
Here we have used $ {\tilde n}^2 = 0$ and the symmetry relation
$K_{kl}(\underline{\alpha}) =K_{lk}(\underline{\alpha})$ which is
proved in Appendix C\@. After summation over all allowed  graphs
we  obtain  the 1PI vertex function
\begin{eqnarray}
\label{B14}
G(\underline{p}|O(\underline{\kappa};\tilde n)) = \sum_\Gamma G_\Gamma
(\underline{p}|O(\underline{\kappa};\tilde n)).
\end{eqnarray}

Let us now turn to the support properties of the anomalous dimension with
respect to the variables $\underline{\kappa}$ and $\underline{\kappa '}$. We
use the fact that the partly unrenormalized 1PI vertex function
$G(\underline{p}|O(\underline{\kappa};\tilde n))$, the renormalized functions
$RG(\underline{p}|O(\underline{\kappa};\tilde n))$ and the anomalous dimension
$\gamma (\underline{\kappa } ,\underline{\kappa ' } )$
have the same support. So we study the support of the vertex function
(\ref{B14}). Its support restrictions follow from the  properties  of
the $\kappa$-dependent part at $\xi_1=...=\xi_L = 0  $ in the exponential:
\begin{eqnarray}
&&\sum_{i,j=1}^2 \kappa_i B_{ij} (\tilde np_j) = 
\nonumber\\
&&\qquad\quad ={\kappa_+\over 2}\left[(B_{11}+B_{12}+B_{21}+B_{22})(\tilde
np_+)
        -(B_{11}-B_{12}+B_{21}-B_{22})(\tilde np_-)\right]
\nonumber\\
&&\qquad\qquad -{\kappa_-\over 2}\left[(B_{11}+B_{12}-B_{21}-B_{22})
     (\tilde np_+)-(B_{11}-B_{12}-B_{21}+B_{22}) (\tilde np_-)\right].
\nonumber\\
\end{eqnarray}
In Appendix C we prove
\begin{eqnarray}
\label{B16}
       B_{11}+B_{21}=1,\quad B_{12}+B_{22}=1,\quad 0\leq B_{ij}\leq 1,
                       \qquad K_{1k}+K_{2k} = 0, \quad K_{kl}=K_{lk},
\nonumber\\
\end{eqnarray}
which suggests the definitions
\begin{eqnarray}
\label{B17}
B_+ = B_{11} - B_{22}, \quad B_- = 1 - B_{11} -B_{22}, \quad
|B_+|  \leq 1, \quad |B_-| \leq  1, \quad  |B_+ \pm B_-| \leq 1.
\nonumber\\
\end{eqnarray}
By taking into account these relations we obtain
\begin{eqnarray}
\sum_{i,j=1}^2 \kappa_i B_{ij} (\tilde np_j)-
            \sum_{i=1}^2 \sum_{l=1}^L \kappa_i K_{il} (\tilde n\xi_l) =
    \kappa_+ (\tilde np_+)&-&\kappa_-[B_+(\tilde np_+) +B_- (\tilde np_-)]
\nonumber\\
&+& \sum_{l=1}^L  \kappa_- K_{1l}(\tilde n\xi_l)
\end{eqnarray}
so that the expression for the 1PI vertex function takes the form
\begin{eqnarray}
            G(\underline{p}|O(\underline{\kappa};\tilde n))&=&
                    \sum_\Gamma \int_0^\infty d\underline{\alpha}
                              \prod_{l=1}^L P_l \left({1\over 2i}
                              {\partial\over\partial\xi_l}\right)
                              e^{-i\alpha_l (m_l - i\rho )}
      f_\Gamma (\underline{\alpha};\underline{p};\underline{\xi}) 
\nonumber\\
                 &&\times\exp i\left \{\kappa_+ (\tilde np_+) -
                   \kappa_-[B_+(\tilde np_+) +B_-(\tilde np_-)] +
                   \sum_{l=1}^L \kappa_- K_{1l} (\tilde n\xi_l)
         \right\}|\scriptstyle \underline{\xi}=0 \displaystyle.
\nonumber\\
\end{eqnarray}
The differentiation with respect to $\xi$ generates the scalar products 
$p_ip_j$,  $\kappa_-(\tilde np_+)$  as well as $\kappa_- (\tilde np_-$) 
so that we are able to write
\begin{eqnarray}
G(\underline{p}|O(\underline{\kappa};\tilde n)) = \sum_\Gamma \int_0^\infty
d\underline{\alpha} g_\Gamma(\underline{\alpha};p_ip_j;\kappa_-
\tilde n\underline{p})  e^{i\kappa_+  (\tilde  np_+)  -i\kappa_-[B_+(\tilde
np_+) +iB_- (\tilde    np_-)]},
\end{eqnarray}
where
\begin{eqnarray}
g_\Gamma(\underline{\alpha};p_ip_j;\kappa_-\tilde n \underline{p})&=&
\prod_{l=1}^L P_l \left({1\over 2i}{\partial\over\partial\xi_l}\right)
                 e^{-i\alpha_l(m_l-i\rho )}
    f_\Gamma (\underline{\alpha}; \underline{p};\underline{\xi})
\nonumber\\
              &&\times e^{i\left\{\kappa_+(\tilde np_+)
                -\kappa_-[B_+(\tilde np_+) +B_- (\tilde np_-)] +
                 \sum_{l=1}^L \kappa_- K_{1l} (\tilde n\xi_l)\right\}}
|\scriptstyle \underline{\xi}=0 \displaystyle.
\nonumber\\
\end{eqnarray}
According to the definition of the anomalous dimension (\ref{B3}), (\ref{B4})
and (\ref{B7}), we perform the Fourier transform and find
\begin{eqnarray}
 \gamma (\underline{\kappa } ,\underline{\kappa '}) &=&
        g{\partial\over \partial g} \mbox{res}  \int    {d(\tilde
np_+)\over 2\pi}\int    {d(\tilde np_-)\over 2\pi} R' \sum_\Gamma
       \int_0^\infty d\underline{\alpha} 
 g_\Gamma(\underline{\alpha};p_ip_j;\kappa_-\tilde n \underline{p}) 
\nonumber\\
           &&\times\exp i\left\{\kappa_+ (\tilde np_+)-
             \kappa_-[B_+(\tilde np_+) +B_-(\tilde np_-)]-
             \kappa'_+ (\tilde np_+)-\kappa'_- (\tilde np_-)\right\},
\nonumber\\
\end{eqnarray}
where the $R'$-operation acts  onto  the  coefficient  function 
$g_{\Gamma}$ only. For
convenience, we introduce the variables \cite{Br}
\begin{eqnarray}
w_+={{\kappa_+'-\kappa_+}\over {\kappa_-}},\quad w_-={{\kappa_-'} \over
{\kappa_-}}
\end{eqnarray}
and receive
\begin{eqnarray}
\label{B24}
\gamma (\underline{\kappa } ,\underline{\kappa '})=\kappa_-^{-2}g {\partial
\over \partial g} \mbox{res} R' \sum_\Gamma \int_0^\infty
d\underline{\alpha}  g_\Gamma \left(\underline{\alpha};p_ip_j;
i{\partial \over {\partial w_+}}, i{\partial  \over  {\partial
w_-}}\right)  \delta(B_+ + w_+) \delta(B_-+w_-).
\nonumber\\
\end{eqnarray}
Because of the restrictions (\ref{B17}) of the variables $B_+$ and $B_-$ we
conclude
\begin{eqnarray}
\label{BSR}
\gamma (w_+,w_-)&\neq& 0\qquad \hbox{for}\quad |w_\pm |\leq 1,\mbox{} |w_+\pm
w_-|
\leq 1 \\
\gamma  (w_+,w_-)&=&\kappa_-^2\gamma  (\underline{\kappa},\underline{\kappa
'}).
\end{eqnarray}
Note that the introduction of $\gamma (w_+,w_-)$ was suggested by the
special variable dependence of Eq. (\ref{B24}).

Let us turn back now to QCD in axial gauge. As already mentioned,  we have
to apply a very complicated gluon propagator. It is in the spirit of the
foregoing considerations to use the following $\alpha$-representation
\begin{eqnarray}
\label{B27}
D_{ab}^{\mu \nu }=-\delta_{ab}\int_0^\infty d\alpha 
      \int_0^\infty d\beta && \Bigg[g^{\mu \nu} +
       {i \over 2} \left( \tilde n^\mu {\partial \over \partial 2i \xi_\nu}+
          \tilde n^\nu {\partial \over \partial 2i \xi_\mu} \right) 
\nonumber\\
&&\times\left(
    e^{i\left\{\beta{\tilde n{\partial\over\partial 2i\xi}+i\tau}\right\}}-
    e^{-i\left\{\beta{\tilde n{\partial\over\partial 2i\xi}-i\tau}\right\}}
\right)\Bigg] 
e^{i\alpha (k^2+i\rho)+i2\xi k}|\scriptstyle \xi=0 \displaystyle .
\nonumber\\
\end{eqnarray}
If we insert this propagator instead of the usual one
into the
$\alpha$-representation of a diagram, then all investigations can be
performed in a similar manner as before. However, it is more instructive
to use instead of the $(\alpha ,\beta )$-representation (\ref{B27}) the
formal expression
\begin{eqnarray}
D_{ab}^{\mu\nu}=-\delta_{ab} \int_0^\infty d\alpha
                                   && \left[g^{\mu\nu} -
      {1\over 2}\left(\tilde n^\mu {\partial\over\partial 2i\xi_\nu}
         +\tilde n^\nu{\partial\over\partial 2i\xi_\mu}\right)
       \left({1 \over {\tilde n{\partial\over\partial 2i\xi}+i\tau}}+
    {1\over {\tilde n{\partial\over\partial 2i\xi}-i\tau}}\right)\right]
\nonumber\\
       &&\qquad\times  e^{i\alpha (k^2+i\rho)+i2\xi k}.
\end{eqnarray}
By writing now the  $\alpha$-representation of a graph, we have
to include also the pre-factors of the gluon propagators
\begin{eqnarray} g^{\mu \nu  } - {1 \over 2} \left( \tilde  n^\mu
{\partial  \over
\partial 2i \xi_\nu} + \tilde n^\nu  {\partial  \over  \partial  2i  \xi_\mu}
\right) \left({1 \over {\tilde n{\partial \over \partial 2i \xi}+i\tau}}+{1
\over {\tilde n{\partial \over  \partial  2i  \xi}-i\tau}}\right)  
\end{eqnarray}
into the original product of the polynomials, so that instead of a
polynomial there appears finally a rational function in the
$\xi$-derivative. This may introduce new singularities. However, the
expression in the exponential which is responsible for the support
properties of the anomalous dimensions remains  unchanged.   This
establishes   the   general   support restrictions (\ref{BSR}) for QCD too.

Of course, for special operators there exist additional restrictions. The
operator (\ref{B1})  transforms  under  charge  conjugation  $C$  in  the
following  way (see Appendix D)
\begin{eqnarray}
\label{B30}
C\left(O^a(\kappa_+,\kappa_-;\tilde n) \right)C^+ = \mp O^a
(\kappa_+,-\kappa_-;\tilde n).
\end{eqnarray}
If we apply the charge conjugation to the renormalization group equation
(\ref{BRGE}) then we get
\begin{eqnarray}
\mu {d \over d\mu}O^a (\kappa_+,-\kappa_-;\tilde n)=\int
d^2\underline{\kappa '} \left( \gamma (\underline{\kappa },
\underline{\kappa '}) - 2\gamma_\psi \delta^{(2)} (\underline{\kappa }-
\underline{\kappa ' }) \right) O^a (\kappa'_+,-\kappa'_-;\tilde n) .
\end{eqnarray}
The substitution $\underline{\kappa} \rightarrow -\underline{\kappa},  \
\underline{\kappa '} \rightarrow -\underline{\kappa '}$
and the comparison with the definition of the operator (\ref{B1}) and 
Eq. (\ref{BRGE}) leads to 
$\gamma(\kappa_+,\kappa_-,\kappa'_+,\kappa'_-)
=\gamma(\kappa _+,-\kappa_-,\kappa'_+,-\kappa'_-)$, 
which is equivalent to
\begin{eqnarray}
\label{BCC}
\gamma (w_+,w_-) = \gamma (-w_+,w_-).
\end{eqnarray}

\section{Properties of the Coefficients \\
of the $\alpha$-Representation}
 
In this Appendix we  prove the relations (\ref{B16})\@. We study the
coefficients $D(\underline{\alpha})$, $B_{li}(\underline{\alpha})$ and
$K_{kl}(\underline{\alpha})$ of the $\alpha$-representation of 1PI-Feynman
graphs with  an operator insertion. We assume that the graph contains $V$
vertices (the vertex corresponding to the operator insertion is the first
vertex), $N$ external and $L$ internal lines. To each internal line there
corresponds one $\alpha$ parameter.
According to the momentum flow we define  an orientation  of  the  diagram.
Instead  of  the  coefficients $B_{li}$ and $K_{kl}$ appearing  in    the
$\alpha$-representation we investigate the coefficients
\begin{eqnarray}
b_{li}(\underline{\alpha})=B_{li}(\underline{\alpha}) D(\underline{\alpha}),
\qquad
k_{kl}(\underline{\alpha})=K_{kl}(\underline{\alpha})D(\underline{\alpha}),
\end{eqnarray}
which are homogeneous polynomials  with  respect  to  all $\alpha$ and linear
functions of each $\alpha$ \cite{Za}. They are given  by  topological
formulas.
$D$ is the so-called chord set product sum
\begin{eqnarray}
D(\underline{\alpha})=\sum_{T\in \{T\}}\left(\prod_{l\notin T}\alpha_l
\right).
\end{eqnarray}
Here, $\{T\}$ denotes a set of trees. A tree of the 1PI-graph is generated
by cutting $(L-V)$ lines of the graph so that a connected  graph  remains
which does not contain loops. The degree of homogeneity of $D$ is $L-V$. We
note further definitions. If we consider a special line $l$, then a tree
containing  this line is denoted by $T_l$. If we cut this line, then we divide
the considered tree into two half-trees corresponding to two simple
connected diagrams. The half-tree lying in the direction of the orientation
of the line $l$ is denoted by $T_l^+$.

After these explanations we write down the definition of the coefficients
$b_{li} (\underline{\alpha })$ with the degree of homogeneity $(L-V)$
\begin{eqnarray}
b_{li}(\underline{\alpha})=\sum_{T_l\in\{T\}\atop i\in T_l^+}\left(
\prod_{k \notin T} \alpha_k \right).
\end{eqnarray}
Here, the sum runs over all trees containing the line $l$ whereas the
considered half-tree has to contain the external vertex $i$. The set of
these trees is denoted by $ \{ T_l\}_{|i \in T_l^+}$. Since $\{
T_l\}_{|i \in  T_l^+ } \subset  \{ T\}$ and $\alpha_k \geq  0$ for each
$k$, it follows from the definition of $D(\underline{\alpha })$ and
$b_{li}(\underline{\alpha })$ that
\begin{eqnarray}
0\leq B_{li}(\underline{\alpha})={b_{li}(\underline{\alpha})\over
D(\underline{\alpha })} \leq  1,
\end{eqnarray}
which is the first relation (\ref{B16}).

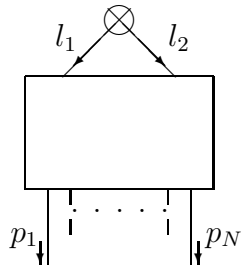
\begin{figure}[h]
\unitlength1cm
\begin{picture}(3.5,4)(-2,-0.5)
\put(1.5,2.5){\line(1,1){0.9}} \put(3,2.5){\line(-1,1){0.9}}
\put(2.25,3.25){\circle{0.4}}
\put(1,1){\line(1,0){2.5}}
\put(1,2.5){\line(1,0){2.5}}
\put(1,1){\line(0,1){1.5}}
\put(3.5,1){\line(0,1){1.5}}
\put(1.3,0){\line(0,1){1}}
\put(1.6,0.8){\line(0,1){0.2}}  \put(1.6,0.4){\line(0,1){0.2}}
\put(1.6,0.7){. . . . .}
\put(2.9,0.8){\line(0,1){0.2}}  \put(2.9,0.4){\line(0,1){0.2}}
\put(3.2,0){\line(0,1){1}}
 
\put(2,3){\vector(-1,-1){0.36}}  \put(1.4,2.9){$l_1$}
\put(2.5,3){\vector(1,-1){0.36}}  \put(2.9,2.9){$l_2$}
\put(1.2,0.3){\vector(0,-1){0.3}}  \put(0.8,0.3){$p_1$}
\put(3.3,0.3){\vector(0,-1){0.3}}  \put(3.4,0.3){$p_N$}
\end{picture}

\caption[FIG.4]
{Topology of the 1PI-Feynman graphs  for  the light-cone
operators in light-cone gauge. The  box  symbolizes  the
connected $(N+2)$-point function. The lines $l_1$ and $l_2$ are  contained
in a loop $c$.\\ }
\label{FIG4}
\end{figure}

To prove the identity
\begin{eqnarray}
\label{C5}
B_{1i} (\underline{\alpha})+B_{2i}(\underline{\alpha})=
{b_{1i}(\underline{\alpha})+b_{2i}(\underline{\alpha})\over
D(\underline{\alpha })} = 1, \quad i = 1,\ldots ,V,
\end{eqnarray}
we have to show that $\{ T\}$ is just the disjoint union of the sets
$\{ T_1\}_{|i \in  T_1^+}$ and $\{ T_2 \}_{|i \in T_2^+}$.
Since $T\in \{ T\}$ is connected, it has to contain the line
$l_1$ or $l_2$ (see Fig. \ref{FIG4}). Therefore $T$ is contained in the sets
$\{ T_1\}_{|i \in  T_1^+ }$ or $\{ T_2 \}_{|i \in  T_2^+ }$.
Moreover $\{ T\}=\{ T_1\}_{|i\in T_1^+}\cup \{ T_2 \}_{|i\in T_2^+}$.
It remains to show that the sets $\{ T_1\}_{|i \in  T_1^+ }$ and $\{ T_2
\}_{|i \in  T_2^+}$ are disjoint. If the vertex $i$ is contained
in a tree $T$ belonging to the sets $\{ T_1\}_{|i \in  T_1^+ }$ and
$\{ T_2 \}_{|i \in  T_2^+ }$ then it is possible to find a path from the
vertex $i$ over the line $l_1$ to the operator vertex and over the line $l_2$
back to the vertex $i$. But this is not possible because $T$ cannot contain a
loop. Therefore the set $\{ T\}$ is the disjoint union of the sets
$\{ T_1\}_{|i\in T_1^+}$ and $\{ T_2 \}_{|i\in T_2^+}$ which proves  Eq. 
(\ref{C5}).

For the investigation of the properties of the $K_{kl} (\underline{\alpha
})$ we start from the definition
\begin{eqnarray}
\label{C6}
k(\underline{\alpha},\underline{\xi})=\sum_{k,l=1}^L\xi_k k_{kl}
(\underline{\alpha})\xi_l=\sum_{T_c\in \{ T_c \}}\left(\prod_{l\notin T_c}
\alpha_l \right)\left( \sum_{j \in c}(\pm \xi_j )\right)^2 .
\end{eqnarray}
Here, $T_c$ denotes a pseudo tree (this means a tree to which we add a line
in such a way that the new structure contains a loop $c$)\@. The sum
$\left(\sum_{j\in c}(\pm\xi_j)\right)^2$ runs over all
lines $j$ which form the loop $c$.
When the line $j \in c$ points along the orientation of the loop c
then we have to write a plus sign, in the opposite case a minus sign.
Obviously the differentiation with respect to $\xi$ provides the symmetry
relation $K_{kl} (\underline{\alpha }) = K_{lk}(\underline{\alpha })$.
Furthermore, the $\xi_1$- and $\xi_2$-dependent part of relation (\ref{C6})
is  determined by those $T_c$ in which $l_1$ and $l_2$ are contained in the
loop $c$. Then, because of our conventions, the lines $l_1$ and $l_2$
have different orientations so that we can write
\begin{eqnarray}
\label{C7}
k(\underline{\alpha},\underline{\xi})=k(\underline{\alpha},\xi_1-\xi_2,
\xi_3,\ldots ,\xi_L).
\end{eqnarray}
The desired relation $K_{1l}
(\underline{\alpha }) = -K_{2l} (\underline{\alpha })$
is then a simple consequence of Eq. (\ref{C7}).

\section{Charge Symmetry}
 
We  prove the following behaviour of the renormalized operators
$O^a(\kappa_+ ,\kappa_- )$ under charge ($C$)-conjugation:
\begin{eqnarray}
\label{D1}
CO^a(\kappa_+,\kappa_-)e^{iS_I}C^+=\mp O^a(\kappa_+,-\kappa_-).
\end{eqnarray}
Let us begin with  the  C-conjugation  of  the  elementary fields.
\begin{eqnarray}
C\psi_\alpha^i(x)C^+&=&\sum_{\beta}c_{\alpha\beta}\overline{\psi}_\beta^i(x),
\quad CA^\mu (x)C^+=-A^\mu (x),
\nonumber\\
C\overline{\psi}_\alpha^i(x)C^+&=&\sum_{\beta}\psi_\beta^i(x)
\stackrel{T_{-1}}{c_{\beta \alpha}},
\end{eqnarray}
where $i$ and $\alpha$, $\beta$ denote the flavour and the spinor indices
of the quark field $\psi$. The transposed matrix of $c:=\{ c_{\alpha \beta
}\}$ is denoted by $\stackrel{T}{c}$. Using  the relation
\begin{eqnarray}
\stackrel{T_{-1}}{c_{\mbox{ }\mbox{}}}\gamma_\mu c=\stackrel{T}{\gamma}_\mu
\end{eqnarray}
we find for the charge conjugation of the operator $:\!\overline{\psi}
(x)\gamma_\mu \lambda^a U(x,y) \psi (y)\! :$
\begin{eqnarray}
\label{D4}
C:\!\overline{\psi}(x)\gamma_\mu \lambda^a U(x,y)\psi(y)\!:C^+&=&
\sum_{\alpha\beta}:\!\psi_\alpha^i (x)(\stackrel{T_{-1}}{c_{\mbox{ }
\mbox{  }}}\!\gamma_\mu c)_{\alpha\beta}\lambda^a_{ij} [CU(x,y)C^+ ]
\overline{\psi}_\beta^i (y)\!: 
\nonumber\\
&=&\sum_{\alpha \beta}:\!\psi_\alpha^i(x)(\gamma_\mu )_{\beta\alpha}
\lambda^a_{ij} [CU(x,y)C^+ ] \overline{\psi}_\beta^i\! : 
\nonumber \\
&=& -:\!\overline{\psi}(y)\gamma_\mu\stackrel{T}{\lambda^a} U(y,x)\psi(x)\!:.
\end{eqnarray}
where we have  used the anticommutation rule of the fermionic field as
well as
\begin{eqnarray}
CU(x,y)C^+=P\exp\left\{-ig\int_y^x dz_\mu CA^\mu (z)C^+\right\}=
U(y,x).
\end{eqnarray}
The hermitean generators $\lambda^a$  satisfy the relation
$\stackrel{T}{\lambda^{a}}=\pm \lambda^a$.
Consequently, setting in Eq. (\ref{D4}) $x=\kappa_1 \tilde n$, $y=\kappa_2
\tilde n$, we find
 \begin{eqnarray}
C:\!\overline{\psi}(\kappa_1\tilde n)\gamma_\mu\lambda^aU(\kappa_1\tilde n,
\kappa_2\tilde n)\psi (\kappa_2\tilde n)\! :C^+=\mp :\!\overline{\psi}
(\kappa_2 \tilde n)\gamma_\mu \lambda^a U(\kappa_2 \tilde n,\kappa_1 \tilde
n) \psi (\kappa_1 \tilde n)\!:
\nonumber\\
\end{eqnarray}
or
\begin{eqnarray}
CO^a(\kappa_+,\kappa_-)C^+=\mp O^a(\kappa_+,-\kappa_-), \quad
\kappa_\pm = (\kappa_2 \pm \kappa_1)/2.
\end{eqnarray}
In QCD $C$-invariance leads to relation (\ref{D1}).

\section{Limits}

After some algebra we obtain the following results for the structures
appearing in the BL-kernels:
\begin{mathletters}
\begin{eqnarray}
\label{E1a}     
             \mbox{LIM} {x\over y} = 1-z,
\end{eqnarray}
\begin{eqnarray}
\label{E1b}
         \mbox{LIM} F(x,y)={1+z^2\over 1-z},
\end{eqnarray}
\begin{eqnarray}
\label{E1c}
\mbox{LIM} F(x,y)\ln\left({x\over y}\right)= {1+z^2\over
1-z}\ln(z)+1-z,
\end{eqnarray}
\begin{eqnarray}
\label{E1d}
         \mbox{LIM} {x\over 2\overline{y}}\ln\left({x\over y}\right)=
             -{1\over 2}\left((1+z)\ln(z)+1-z\right),
\end{eqnarray}
\begin{eqnarray}
\label{E1e}
        \mbox{LIM}\left(F(x,y)+{x\over 2\overline{y}}\right)
                     \ln^2\left({x\over y}\right)=
              {1+3z^2\over 2(1-z)}\ln^2(z)+(1-z)\ln(z),
\end{eqnarray}
\begin{eqnarray}
\label{E1f}
         \mbox{LIM} [F(x,y)-\overline{F}(x,y)] \ln\left({x\over y}\right)
             \ln\left(1-{{x}\over {y}}\right) =
                  2{1+z^2\over 1-z}\ln(z)\ln(1-z),
\end{eqnarray}
\begin{eqnarray}
\label{E1g}
      \lefteqn{\mbox{LIM}[\overline{F}(x,y)\ln(y)\ln(\overline{x})-
                 F(x,y)\hbox{Li}_2(\overline{y})+
              \overline{F}(x,y)\hbox{Li}_2(\overline{x})]} 
\nonumber\\
    &&=-{1+z^2\over 1-z}\left({\pi^2\over 6}-{1\over 2}\ln^2(z)\right)
                  +(1+z)\ln(z)+2(1-z),
\end{eqnarray}
\begin{eqnarray}
\label{E1h}
\lefteqn{\mbox{LIM}F(x,y)\left[2\hbox{Li}_2\left(1-{x\over y}\right)+
            \ln^2(y)-2\ln(x)\ln(y)+
          2\hbox{Li}_2(\overline{y})-2\hbox{Li}_2 (x)
              +\ln^2(\overline{y})\right.}
\nonumber\\
          &&\hspace{11cm}+\left.
2\hbox{Li}_2\left(1-{\overline{x}\over\overline{y}}\right)\right]
\nonumber\\
&&=2{1+|z|^2\over 1+|z|}\left(\hbox{Li}_2\left({|z|\over 1+|z|}\right)
    -\hbox{Li}_2\left({1\over 1+|z|}\right)+{1\over 2}\ln^2 (|z|)
     -\ln(|z|)\ln(1+|z|)\right)
\nonumber\\
          &&\hspace{6cm}\mbox{}+2(1+|z|)\ln(|z|)+4(1-|z|),
\end{eqnarray}
\begin{eqnarray}
\label{E1i}
   \mbox{LIM}\left[{x\over
2(\overline{y})}\ln(x)(1+\ln(x)-2\ln(\overline{x}))+
           \left\{{x\leftrightarrow\overline{x}\atop
          y\leftrightarrow\overline{x}}\right\} \right]=0.
\end{eqnarray}
\end{mathletters}
Whereas the determination of most of these expressions is very simple, those
containing the Li$_2$-functions are not so easy to handle.
 
To determine the limit of $G(x,y)$ in Eq. (\ref{Gxy}) we have to prove Eqs.
 (\ref{E1g}) and (\ref{E1h}). First,  the expression
\begin{eqnarray}
         \lim_{\tau\to 0}\tau^{-1}\left(2\overline{F}[\ln(y)
     \ln(\overline{x})+\hbox{Li}_2(y)+\hbox{Li}_2(\overline{x})]-
                   2F[\ln(\overline{y})\ln(x)+
      \hbox{Li}_2(\overline{y})+\hbox{Li}_2(x)]\right).
\nonumber\\
\end{eqnarray}
 may be simplified. Taking into account
$\hbox{Li}_2(x)+\hbox{Li}_2(1-x)=-\ln(x)\ln(1-x)+\pi^2/6$, we get
\begin{eqnarray}
       \ln(y)\ln(\overline{x})+\hbox{Li}_2(y)+\hbox{Li}_2(\overline{x})
                          &=&{\pi^2\over 6}+\hbox{Li}_2(\overline{x})-
                             \hbox{Li}_2(\overline{y})-
        \ln(y)\ln\left({\overline{x}\over\overline{y}}\right),
\nonumber\\
       \ln(\overline{y})\ln(x)+\hbox{Li}_2(\overline{y})+\hbox{Li}_2(x)
                          &=&{\pi^2\over 6}+\hbox{Li}_2(\overline{y})-
                              \hbox{Li}_2(\overline{x})-
         \ln(x)\ln\left({\overline{x}\over\overline{y}}\right).
\end{eqnarray}
From this, together with  \cite{Le}
\begin{eqnarray}
\label{E4} 
    \hbox{Li}_2(\tau )&=&\tau+{1\over 4} \tau^2  +\cdots 
\nonumber\\
     \hbox{Li}_2\left(1-{z\over\tau}\right)&=&
    -\hbox{Li}_2\left({\tau\over\tau -z}\right)-
     {1\over 2}\ln^2\left({z-\tau\over\tau}\right)-{\pi^2\over 6}\nonumber\\
     &=&{\tau\over z}-{1\over 2}\ln^2\left({z-\tau\over\tau}\right)
            -{\pi^2\over 6}+ \cdots 
\end{eqnarray}
and the corresponding expansion of $F$ and $\overline{F}$, we obtain the
result (\ref{E1g}).
 
Consider now the term within the second bracket of (\ref{Gxy}),
\begin{eqnarray}
      &&\theta\left({\overline{x}\over y}\right)
              \theta\left(1-{\overline{x}\over y}\right)\mbox{sign}(y)
       \left[2(F-\overline{F})\hbox{Li}_2\left(1-{x\over y}\right)+
           (F-\overline{F})\ln^2 (y)-2F\ln(x)\ln(y)\right.
\nonumber\\
&&\hspace{6cm}\mbox{}+2F\hbox{Li}_2(\overline{y})
           -2F\hbox{Li}_2(x)\bigg]
        +\left\{{x \leftrightarrow\overline{x} \atop y
              \leftrightarrow\overline{y}}\right\}.
\end{eqnarray}
According to our procedure we have to calculate
\begin{eqnarray}
E(z)&=&\lim_{\tau\to 0}\tau^{-1}\left[F(x,y)
\left(2\hbox{Li}_2\left(1-{x\over y}\right)+
\ln^2(y)-2\ln(x)\ln(y)+2\hbox{Li}_2(\overline{y})-
2\hbox{Li}_2(x)\right.\right.
\nonumber\\
&&\hspace{2.5cm}+\left.\left.\ln^2(\overline{y}) +
2\hbox{Li}_2\left(1-{\overline{x}\over\overline{y}}\right)\right)
-\left\{{x\leftrightarrow\overline{x}\atop
y\leftrightarrow \overline{y}}\right\}
\right]|\scriptstyle x={z\over\tau},y={1\over\tau}\displaystyle .
\end{eqnarray}
Here, we apply
$\hbox{Li}_2 (x)+\hbox{Li}_2(1/x)=-{1\over 2}\ln^2(-x)-\pi^2/6$
to express
\begin{eqnarray}
           \hbox{Li}_2\left(1-{x\over y}\right)&=&
     -\hbox{Li}_2\left({x\over y}\right)-\ln\left({x\over y}\right)
           \ln\left(1-{x\over y}\right)+{\pi^2 \over 6},
\nonumber\\
       \hbox{Li}_2\left(1-{\overline{x}\over\overline{y}}\right)&=&
        -\hbox{Li}_2\left({\overline{y}\over x-y}\right)
   -{1\over 2} \ln^2\left({y-x\over\overline{y}}\right)-{\pi^2 \over 6}
\end{eqnarray}
and the relation
$\hbox{Li}_2(1-x)+\hbox{Li}_2\left(1-{1\over x}\right)=-{1\over 2}\ln^2(x)$
to get
\begin{eqnarray}
               \hbox{Li}_2\left(1-{x\over y}\right)=
              -\hbox{Li}_2\left({x\over x-y}\right)-
             {1\over 2}\ln^2\left(1-{x\over y}\right).
\end{eqnarray}
Then we obtain
\begin{eqnarray}
&&E(z)= 
\nonumber\\
&&\mbox{}=\lim_{\tau\to 0}\tau^{-1}\left[2F\left(\hbox{Li}_2(\overline{y})-
        \hbox{Li}_2(x)-\hbox{Li}_2\left({\overline{y}\over x-y}\right)+
        \hbox{Li}_2\left({x\over x-y}\right)+
        \ln(y-x)\ln\left({\overline{y}\over x}\right)\right)\right.
\nonumber\\
       &&\mbox{}-2\overline{F}\left.\left(\hbox{Li}_2(x)-
        \hbox{Li}_2(\overline{y})-
   \hbox{Li}_2\left({\overline{y}\over x-y}\right)+
        \hbox{Li}_2\left({x\over x-y}\right)+
        \ln\left({y-x\over y\overline{x}}\right)
        \ln\left({\overline{y}\over x}\right)
        \right)\right]|\scriptstyle
x={z\over\tau},y={1\over\tau}\displaystyle.
\nonumber\\
\end{eqnarray}
From this transformed expression we are  able  to  determine (\ref{E1h})
using (\ref{E4}) together with
\begin{eqnarray}
\hbox{Li}_2(x+\epsilon )=\hbox{Li}_2(x)-\epsilon\ln(1-x)/x+\cdots.
\end{eqnarray}





\end{document}